UNCORRECTED PREPRINT

# A Perspective from Extinct Radionuclides on a Young Stellar Object: The Sun and Its Accretion Disk

Nicolas Dauphas,[1] Marc Chaussidon[2]

[1]Origins Laboratory, Department of the Geophysical Sciences and Enrico Fermi Institute, Chicago, Illinois 60637; email: dauphas@uchicago.edu

[2]Centre de Recherches Pétrographiques et Géochimiques (CRPG)-Nancy Université-CNRS, UPR 2300, 54501 Vandoeuvre-lès-Nancy, France

■ **Abstract** Meteorites, which are remnants of solar system formation, provide a direct glimpse into the dynamics and evolution of a young stellar object (YSO), namely our Sun. Much of our knowledge about the astrophysical context of the birth of the Sun, the chronology of planetary growth from micrometer-sized dust to terrestrial planets, and the activity of the young Sun comes from the study of extinct radionuclides such as $^{26}$Al ($t_{1/2}$ = 0.717 Myr). Here we review how the signatures of extinct radionuclides (short-lived isotopes that were present when the solar system formed and that have now decayed below detection level) in planetary materials influence the current paradigm of solar system formation. Particular attention is given to tying meteorite measurements to remote astronomical observations of YSOs and modeling efforts. Some extinct radionuclides were inherited from the long-term chemical evolution of the Galaxy, others were injected into the solar system by a nearby supernova, and some were produced by particle irradiation from the T-Tauri Sun. The chronology inferred from extinct radionuclides reveals that dust agglomeration to form centimeter-sized particles in the inner part of the disk was very rapid (<50 kyr), planetesimal formation started early and spanned several million years, planetary embryos (possibly like Mars) were formed in a few million years, and terrestrial planets (like Earth) completed their growths several tens of million years after the birth of the Sun.

# 1. INTRODUCTION

In the past decade, considerable effort has been directed toward integrating astronomical observations of young stellar objects (YSOs), models of solar system formation, and laboratory studies of meteorites and comets to yield a general understanding of the birth of our Sun and the formation of the planets (**Figure 1**; see also recent reviews in Davis 2005, Krot et al. 2005c, Reipurth et al. 2007). The Sun was born in a molecular cloud, probably nearby other newborn stars (Adams 2010, Lada & Lada 2003). During collapse of the solar system's parent molecular cloud core, conservation of angular momentum led to the formation of a protoplanetary disk (or proplyd), through which material accreted by the Sun was channeled. In the inner part of the disk, most material was condensed from cooling gas of near-solar composition (Grossman 1972, Grossman & Larimer 1974). The dust particles thus created, as well as those inherited from the parent molecular cloud (0.1--10 μm), are thought to have aggregated by Brownian motion, differential vertical settling, and turbulence (Blum & Wurm 2008, Dullemond & Dominik 2005, Weidenschilling 1980). In the size range of 10–100 cm, dust agglomerates experienced a headwind from slower-rotating gas, which led them to spiral toward the protostar (Adachi et al. 1976, Weidenschilling 1977). Possibly helped by gravitational instability in a turbulent disk (Cuzzi et al. 2008, Johansen et al. 2007), dust agglomerates coalesced into rapidly growing planetesimals (1–1,000 km in diameter) that were largely immune to the effects of a headwind known as gas drag. Models suggest that the subsequent growth of planetesimals led to the formation of several Mars-sized planetary embryos (1,000–5,000 km) regularly interspaced in heliocentric distances (Goldreich et al. 2004, Kokubo & Ida 1998, Safronov 1969, Wetherill & Stewart 1989). Collision of these embryos is believed to have led to the formation of terrestrial planets such as Earth (12,742 km) (Chambers & Wetherill 1998). At some point in this evolution, nebular gas not accreted onto the protostar was lost by photoevaporation processes (Alexander 2008, Hollenbach et al. 2000).

< PLEASE INSERT FIGURE 1 HERE>

**Figure 1.** Evolution of matter in the inner part of the solar protoplanetary disk. Solid dust was condensed from nebular gas (Grossman 1972, Grossman & Larimer 1974) and was inherited from the molecular cloud core that formed the Sun (Mathis et al. 1977, Weingartner & Draine 2001). Brownian motion, differential vertical settling, and turbulence led these particles to agglomerate into fractal agglomerates held together by van der Waals forces (Blum & Wurm 2008, Dullemond & Dominik 2005, Weidenschilling 1980). As the relative velocities of dust agglomerates increased, collisions led to compaction and eventually to fragmentation and bouncing. The size threshold above which collisions became disruptive or ineffective is <1 m. The physical process by which dust agglomerates grew into planetesimals is not well understood but could have involved gravitational instability in dust-enhanced regions of a turbulent disk (Cuzzi et al. 2008, Johansen et al. 2007). Orderly, runaway, and oligarchic growth of planetesimals formed several Moon- to Mars-sized embryos at regular heliocentric distances (Goldreich et al. 2004, Kokubo & Ida 1998, Safronov 1969, Wetherill & Stewart 1989); these embryos later collided to form terrestrial planets (Chambers & Wetherill 1998). Abbreviation: ISM, interstellar medium.

In the present context, extinct radionuclides (Table 1) are defined as short-lived nuclides that were present at the time of solar system formation but have since decayed below detection level (McKeegan et al. 2006). These extinct radionuclides have diverse origins. Some, such as $^{146}$Sm ($t_{1/2}$ = 103 Myr), were inherited from the long-term chemical evolution of the Galaxy (Clayton 1988, Nittler & Dauphas 2006). Others, such as $^{26}$Al ($t_{1/2}$ = 0.717 Myr), were produced in one or several nearby stars and were mixed with solar system material shortly before formation of planetary bodies (Cameron & Truran 1977, Meyer & Clayton 2000, Takigawa et al. 2008, Wasserburg et al. 2006). A few, such as $^{10}$Be ($t_{1/2}$ = 1.39 Myr; Chmeleff et al. 2010), were produced by intense particle irradiation from the young Sun (Goswami et al. 2001, Gounelle et al. 2001, Leya et al. 2003, McKeegan et al. 2000). In this review, we cover how the abundance and distribution of extinct radionuclides in meteorites can be used to constrain the astrophysical setting of solar system formation, the efficiency of large-scale mixing in the protoplanetary disk, the intensity of solar flares at solar system birth, and the chronology of planetary growth and differentiation.

<COMP: PLEASE INSERT TABLE 1 HERE>

**Table 1**. Extinct radionuclides in meteorites[a]The ★★★ symbol means that the extinct radionuclide was present when the solar system formed and that its initial abundance has been precisely determined in some primitive components of meteorites. [b]★★ means that there is convincing evidence for the presence of the extinct radionuclide but that its initial abundance is uncertain. [c]★ means that there are reports of the extinct radionuclide in meteorites but that the evidence is weak and awaits confirmation.[d]< means that the extinct radionuclide has not been detected (only an upper limit on its initial abundance exists).

All lithic objects that preserve a record of extinct radionuclides derive from meteorites. Meteorites are solid remnants of the formation of the solar system and present a wide variety of characteristics, yet they are amenable to systematic classification (Krot et al. 2005b). One distinguishes, for example, between undifferentiated (unmolten) and differentiated (molten) meteorites. The unmolten meteorites are also termed chondrites because they contain chondrules, which are millimeter- to centimeter-sized, partially crystallized silicate droplets. Coexisting with these chondrules are refractory inclusions (calcium-, aluminum-rich inclusions, or CAIs) that are approximately the same size but differ in their highly refractory chemical compositions. Chrondrules and CAIs are bound together by fine dust known as matrix. Although many meteorites have suffered some mineralogic alteration of their "parent bodies" (asteroids), these constituents represent a good sampling of the dust that was around in the nebula when planetary bodies formed. Differentiated meteorites were once molten and are taken to represent the mantles (achondrites) or the cores (magmatic iron meteorites) of planetesimals or planets.

The past presence of extinct radionuclides in meteorites and other solar system materials can be inferred from the study of their decay products. Thus, the past presence of the first extinct radionuclide discovered, $^{129}$I ($t_{1/2}$ = 15.7 Myr), was revealed through measurement of the abundance of $^{129}$Xe in meteorites (Jeffery & Reynolds 1961). Because of its half-life and the fact that Al can be fractionated easily from Mg by nebular and magmatic processes, $^{26}$Al ($t_{1/2}$ = 0.717 Myr; decays into $^{26}$Mg) is more widely used to establish the chronology of solar system formation. A diagram describing its use is shown in **Figure 2**. If several rocks or minerals formed at the same time (same initial $^{26}$Al/$^{27}$Al ratio) from homogeneous material (same initial $^{26}$Mg/$^{24}$Mg ratio), then Al-rich minerals such as anorthite (CaAl$_2$Si$_2$O$_8$) should develop more radiogenic $^{26}$Mg/$^{24}$Mg ratios compared with those of more magnesian minerals such as spinel (MgAl$_2$O$_4$). Lee et al. (1976, 1977) used these principles to demonstrate that $^{26}$Al was present at the time of formation of the first solids in the solar system at a level of $^{26}$Al/$^{27}$Al ≈ 5 × 10$^{-5}$ (see also Gray & Compston 1974). As demonstrated in this review, short-lived radionuclides such as $^{26}$Al and $^{10}$Be provide a record of the Sun as a YSO and of the evolution of its accretion disk (**Figure 3**).

< PLEASE INSERT FIGURE 2 HERE>

**Figure 2** Principles behind the use of the $^{26}$Al-$^{26}$Mg decay system in meteorites ($t_{1/2}$ = 0.717 Myr). Until an object forms, the $^{26}$Mg/$^{24}$Mg ratio evolves in a reservoir of chondritic $^{27}$Al/$^{24}$Mg ratio (<u>ch</u>ondritic <u>u</u>niform <u>r</u>eservoir, or CHUR). When object 0 forms at $t_0$, partitioning of Al and Mg between several phases fractionates the $^{27}$Al/$^{24}$Mg ratio. After decay of $^{26}$Al in that object, the $^{26}$Mg/$^{24}$Mg ratio must correlate with the $^{27}$Al/$^{24}$Mg ratio according to the equation (Lee et al. 1976, 1977)
$$\left(^{26}Mg/^{24}Mg\right)_{measured\,at\,present} = \left(^{26}Mg/^{24}Mg\right)_{initial} + \left(^{26}Al/^{27}Al\right)_{initial} \times \left(^{27}Al/^{24}Mg\right)_{measured\,at\,present}$$
$$\left(\frac{^{26}Mg}{^{24}Mg}\right)_{measured\,at\,present} = \left(\frac{^{26}Mg}{^{24}Mg}\right)_{initial} + \left(\frac{^{26}Al}{^{27}Al}\right)_{initial} \times \left(\frac{^{27}Al}{^{24}Mg}\right)_{measured\,at\,present}.$$
The slope of the correlation gives the $^{26}$Al/$^{27}$Al ratio at $t_0$, whereas the intercept gives the $^{26}$Mg/$^{24}$Mg ratio at $t_0$. When object 1 forms at $t_1$, the $^{26}$Mg/$^{24}$Mg ratio in CHUR increases owing to decay of $^{26}$Al. The measured correlation between $^{26}$Mg/$^{24}$Mg and $^{27}$Al/$^{24}$Mg in object 1 should therefore have lower slope (lower initial $^{26}$Al/$^{27}$Al ratio) and higher intercept (higher initial $^{26}$Mg/$^{24}$Mg ratio) than those measured in object 0. The time difference between formation of objects 0 and 1 is given by $t_1 - t_0 = \frac{1}{\lambda_{^{26}Al}} \ln \frac{(^{26}Al/^{27}Al)_{t_0}}{(^{26}Al/^{27}Al)_{t_1}}$. $t_1 - t_0 = (1/\lambda_{^{26}Al}) \times \ln\left[\left(^{26}Al/^{27}Al\right)_{t_0} / \left(^{26}Al/^{27}Al\right)_{t_1}\right]$. In isochron diagrams, $\delta^{26}$Mg* = [($^{26}$Mg/$^{24}$Mg)$_{sample}$/($^{26}$Mg/$^{24}$Mg)$_{CHUR}$ − 1] × 10$^3$ is used in place of the $^{26}$Mg/$^{24}$Mg ratio. This is only a matter of convenience. It is, for example, easier to report that a sample has excess $\delta^{26}$Mg* of 1 ‰ relative to CHUR rather than to report two $^{26}$Mg/$^{24}$Mg isotopic ratios of 0.13946 and 0.13932 for the sample and CHUR, respectively. The relationship between $\delta^{26}$Mg* and the $^{27}$Al/$^{24}$Mg ratio takes the form $\delta^{26}Mg^*_{measured\,at\,present} = \delta^{26}Mg^*_{initial} + \left(\frac{^{26}Al}{^{27}Al}\right)_{Initial} \left(\frac{^{27}Al}{^{24}Mg}\right)_{measured\,at\,present} \left(\frac{^{24}Mg}{^{26}Mg}\right)_{CHUR} \times 10^3$.
$\delta^{26}Mg^*_{measured\,at\,present} = \delta^{26}Mg^*_{initial} + \left(^{26}Al/^{27}Al\right)_{initial} \left(^{27}Al/^{24}Mg\right)_{measured\,at\,present} \left(^{24}Mg/^{26}Mg\right)_{CHUR} \times 10^3$.

< PLEASE INSERT FIGURE 3 HERE>

**Figure 3** $^{26}$Al record of solar system formation anchored to astrophysical observations of nascent stars (adapted from André 2002). (*Left*) The five different stages, from prestellar core to class 0–III stellar sources, are identified from the observed infrared excess relative to a blackbody spectrum. Observations of young stellar objects (YSOs) show that the duration from class 0 to class III is at most ~10 Myr (André 2002, Cieza et al. 2007, Montmerle et al. 2006, Reipurth 2005). (*Right*) $^{26}$Al variations in meteoritic components provide a way to define independently a similar timescale for the young Sun, which can be followed by the variations of the $^{26}Al/^{27}Al$ ratio (the slopes of the $^{26}Al$ isochrons shown in the panels *a–d*). Panel *a* shows an image, although strongly biased, of the isotopic heterogeneity (fossil excesses of $^{26}Al$ ) carried by grains present in the parent cloud. Black dots correspond to a subset of SiC presolar grains from the Orgueil chondrite (Huss et al. 1997). The $^{26}Al$ galactic background (Diehl et al. 2006) is shown by a thick line in panels *a* and *b*. The star in panel *a* corresponds to the possible composition of the accretion disk characterized by a $^{26}Al/^{27}Al$ ratio of $5 \times 10^{-5}$ after injection of Al and Mg isotopes from a nearby supernova. Panel *b* shows time zero for $^{26}Al$ chronology, which can be defined by the formation of calcium-, aluminum-rich inclusions (CAIs) from CV chondrites (a class of CAI-rich carbonaceous chondrites that are named after the Vigarano meteorite). This happened when $^{26}Al/^{27}Al$ was homogenized to $5.2 \times 10^{-5}$ in the inner solar system (Bouvier & Wadhwa 2010, Jacobsen et al. 2008, MacPherson et al. 2010a,b). The presence in these CAIs of short-lived $^{10}Be$ made by irradiation from the young active Sun links this "age zero" to the Sun when it was a class II (active T-Tauri) star (or perhaps even a class I star). This means that, in theory, meteorites might contain older (by up to a few hundred thousand years at most) components. Accretion processes in the disk were coeval with the formation of CAIs and led to the rapid formation of the first generation of planetesimals, as shown by short-lived $^{182}Hf$ in some magmatic iron meteorites. Panel *c* shows the formation times of two chondrules (*open and solid dots*) from the Semarkona chondrite characterized by $^{26}Al/^{27}Al$ ratios of $7.9 \times 10^{-6}$ and $3.4 \times 10^{-6}$ (data from Villeneuve et al. 2009). Chondrules were formed by high-temperature processes in the disk over a time interval of ~1 to 4 Myr after CAIs. Panel *d* shows the melting age of the parent body of angrites, a class of differentiated meteorites. The data define an isochron with a slope corresponding to an initial $^{26}Al/^{27}Al$ ratio of $5.1 \times 10^{-7}$ (Schiller et al. 2010, Spivak-Birndorf et al. 2009), i.e., a melting age of ~5 Myr after CAIs. The angrite parent body must have accreted ~2 Myr earlier. The earlier accretion of the parent bodies of some magmatic iron meteorites cannot be dated with $^{26}Al$. Planetesimals and planetary embryos (the size of Mars or Mars itself) were presumably formed when the Sun was still a T-Tauri star.

## 2. THE ASTROPHYSICAL CONTEXT OF SOLAR SYSTEM FORMATION AND THE EFFICIENCY OF MIXING

Whereas many extinct radionuclides were inherited from the long-term chemical evolution of the Galaxy (Clayton 1988, Nittler & Dauphas 2006), the ones with half-lives of less than a few million years (e.g., $^{26}Al$, $^{41}Ca$, $^{53}Mn$, and $^{60}Fe$) can potentially tie the forming solar system to its immediate astrophysical environment (Adams 2010). Soon after the discovery of $^{26}Al$ (Lee et al. 1976), this nuclide was proposed to be the signature of the explosion of a core-collapse (type II) supernova that would have triggered the collapse of the molecular cloud core that made the Sun (Cameron & Truran 1977). Indeed, its initial abundance ($^{26}Al/^{27}Al = 5.23 \pm 0.13 \times 10^{-5}$; Jacobsen et al. 2008) is approximately one order of magnitude higher than that of the galactic background ($8.4 \pm 2.4 \times 10^{-6}$; Diehl et al. 2006). From the observation of additional extinct radionuclides, this idea was refined to suggest that the forming solar system was polluted by $\sim10^{-4}$ solar mass of freshly nucleosynthesized matter from the outflow of a supernova that exploded ~1 Myr before solar system formation (e.g., Birck & Allègre 1988, Meyer & Clayton 2000, Takigawa et al. 2008, Wasserburg 1987). Another possible scenario is that the solar system was contaminated by the wind from a nearby asymptotic giant branch (AGB) star (see Wasserburg et al. 2006 and references therein). However, the probability of having an AGB star at the time and place of the formation of the solar system is small (Kastner & Myers 1994). Indeed, AGB stars are evolved stars, and their presence in star-forming regions would be coincidental. Conversely, core-collapse supernovae are expected to be present in stellar nurseries. The constraints that can be brought from extinct radionuclides on the putative astrophysical sources are highly dependent on initial solar system abundances. In particular, a high initial $^{60}Fe/^{56}Fe$ ratio would be indicative of a supernova origin for some (Wasserburg et al. 1998). One difficulty with $^{60}Fe$ comes from the fact that it has not been detected in refractory inclusions because of the presence of Ni nucleosynthetic anomalies (Birck & Lugmair 1988, Quitté et al. 2007), and it was identified only in objects formed later. Taken at face value, measurements reported in the literature are contradictory. Measurements of bulk angrites and eucrites (meteorites representing fragments of the mantles of differentiated planetesimals) define an initial $^{60}Fe/^{56}Fe$ ratio of $(5.8 \pm 1.3) \times 10^{-9}$ when back-calculated to the time of CAI formation (Quitté et al. 2010, Tang & Dauphas 2011), whereas chondrules give an initial ratio of $(4.3 \pm 1.4) \times 10^{-7}$ at the time of CAI formation (Mishra et al. 2010, Tachibana et al. 2006). Both values were calculated with the same half-life for $^{60}Fe$ of 2.62 Myr (Rugel et al. 2009). Further work will be required to assess which of these values corresponds to the initial solar system abundance. Until this is done, constraints put on the astrophysical context of solar system formation from $^{60}Fe$ should be taken with a grain of salt. The proposed values for the initial $^{60}Fe/^{56}Fe$ ratio in meteorites encompass the $^{60}Fe/^{56}Fe$ ratio of $1.4 \pm 0.9 \times 10^{-7}$ that can be calculated from analysis of γ-ray emission from the galactic plane ($^{60}Fe/^{26}Al = 0.148 \pm 0.06$; Wang et al. 2007), and an enhancement of this background by a factor of two at 4.5 Gyr is possible (Williams 2010).

Theoretical considerations and observations demonstrate the feasibility of triggering the formation of the solar system by an interstellar shock wave propagating from a supernova on a molecular cloud core (Vanhala & Boss 2002), although a restricted range of conditions (supernova distance, shock-wave velocity) is required (Boss et al. 2008). Alternatively, in case of direct injection from a massive star going supernova near an already formed protoplanetary disk, the disk can survive the flash of ultraviolet radiation from the supernova, the ram pressure (drag force exerted by the ejecta on the disk), and the input of momentum caused by the supernova shock (Ouellette et al. 2005, 2007). In such a scenario, numerical simulations show that the efficiency of injection into the protoplanetary disk should be very low (~1 %) for the gas phase but might be near 100 % for solid grains condensed in the ejecta, which could be the carriers of some extinct radionuclides (Ouellette et al. 2007). The presence, in the right amount, of $^{60}Fe$ and $^{26}Al$ in the solar system's parental molecular cloud can also be viewed as the contribution from numerous supernovae enriching the parcel of gas and dust that made the Sun over a period of ~20 Myr (Gounelle et al. 2009). Thus, there are several solutions to the stellar origin of extinct radionuclides with very short half-lives, but the uncertainties associated with model predictions or with initial abundance determinations from meteorites do not allow us to favor definitely one scenario over another.

Extinct radionuclide abundances in meteorites more usefully constrain mixing processes between the stellar ejecta and the accretion disk. Recently, evidence has been found for a relatively homogeneous distribution (at the ±10 % level) of $^{26}Al$ (Villeneuve et al. 2009). The fundamental difficulty in establishing the homogeneity of $^{26}Al$ is that comparing the $^{26}Al/^{27}Al$ measured in different objects requires that their absolute ages must be determined precisely and independently to correct for rapid $^{26}$Al decay. One way to circumvent this difficulty is to compare the $\delta^{26}$Mg* values of objects that have different $^{26}$Al/$^{27}$Al ratios. ($\delta^{26}$Mg* is the radiogenic excess in the $^{26}$Mg/$^{24}$Mg ratio in parts per thousand, defined as $\delta^{26}$Mg* = [($^{26}$Mg/$^{24}$Mg)$_{sample}$/($^{26}$Mg/$^{24}$Mg)$_{CHUR}$ − 1] × 10$^3$.) In the case of initially homogeneous Mg and Al isotopes in a reservoir of constant Al/Mg ratio (assumed to be solar or chondritic for the accretion disk), their $\delta^{26}$Mg* values should increase proportionally to the decrease of their $^{26}$Al/$^{27}$Al ratios (compare the intercepts and slopes of the present isochrons in **Figure 2**). This effect is small (a ~0.03‰ increase of the $\delta^{26}$Mg* value for complete $^{26}$Al decay from $^{26}$Al/$^{27}$Al = 5.2 × 10$^{-5}$ to 0), but it was demonstrated to be present among chondrules, Earth, and CAIs, thanks to the development of high-precision in situ analyses (Villeneuve et al. 2009). This result constrains Mg and Al isotopes to a homogeneous distribution of ±10 % relative to the average in those three reservoirs (Villeneuve et al. 2009). Similarly, high-precision Ni isotope analyses have shown that early-formed magmatic iron meteorites (see Section 5) have no $^{60}$Ni deficits (and collateral anomalies on $^{58}$Fe and $^{64}$Ni) relative to chondrites, as would be expected if they formed either prior to the injection of $^{60}$Fe or in zones of the accretion disk devoid of $^{60}$Fe (Dauphas et al. 2008). This suggests that the initial $^{60}$Fe was homogeneously distributed at the ±10 % level, but a definitive conclusion on this issue awaits redetermination of the initial $^{60}$Fe/$^{56}$Fe ratio in chondrites. Numerical simulations (Boss 2007) indicate that extinct radionuclides (e.g., $^{26}$Al) injected at the surface of the nebula will be homogenized down to ±10 % in a few thousand years (at least at the scale of the grid resolution used in these simulations), in agreement with the meteoritic record. The results presented here do not prevent the presence of some heterogeneity for some extinct radionuclides at a fine scale. For example, FUN inclusions (FUN is an acronym that refers to the isotope fractionation and unknown nuclear effects measured in these inclusions, which distinguishes them from normal CAIs; Clayton & Mayeda 1977, Wasserburg et al. 1977) and some hibonite inclusions do not show evidence of the presence of $^{26}$Al (MacPherson et al. 1995, Liu et al. 2009), presumably because they formed early, before complete homogenization of this extinct radionuclide occurred. A CAI from a comet yielded no detectable trace of $^{26}$Al (i.e., $^{26}$Al/$^{27}$Al < 10$^{-5}$; Matzel et al. 2010), and further work should tell whether this is due to formation before $^{26}$Al homogenization or formation/resetting after $^{26}$Al decay.

Finally, in a rather counterintuitive way, such rapid mixing processes may also explain the presence of large-scale isotopic heterogeneities in the accretion disk that are known to be present at planetary scales for either major (oxygen; Clayton 2007, Clayton et al. 1973) or trace (molybdenum; Dauphas et al. 2002b) elements. Indeed, such processes could reveal isotopic anomalies from a globally homogeneous medium by aerodynamically sorting the grains in size, if grains of a certain dimension were to carry the isotopic anomalies. This idea was proposed to explain $^{54}$Cr variations measured in bulk planetary materials because the carrier of these anomalies is very fine grained (<100 nm); it would follow the gas and thus be decoupled easily from coarser grains (Dauphas et al. 2010).

## 3. IRRADIATION

Studies of the X-ray activity of low-mass, pre–main sequence stars have revealed that these stars show time-averaged luminosity and power of individual flares that exceed present-day solar values by several orders of magnitude (Feigelson & Montmerle 1999, Feigelson et al. 2002). Although the origin of this strong X-ray activity is incompletely understood, observations of well-characterized T-Tauri stars in the Orion nebular cluster made using the Chandra Orion Ultradeep Project (COUP) indicate that it probably arises from a turbulent dynamo created in the depths of the stellar convection zone, if the T-Tauri stars are fully convective (Preibisch et al. 2005). Alternatively, if the T-Tauri stars are not fully convective, their X-ray activity must result from a solar-like dynamo operating at the base of their convection zones. Energy released during accretion shocks and/or periodical reconnections of the dipolar stellar magnetic field due to differential rotation of the star and the accretion disk, whose inner part is connected to the stellar magnetic field, can heat the gas surrounding the star to X-ray-emitting temperatures (more than $10^7$ K) (Preibisch et al. 2005). Regardless of these uncertainties, it is obvious that circumstellar material around an X-ray-emitting T-Tauri star must be partly photoionized, which should influence accretion and outflow processes as well. COUP observations in the Orion nebula cluster show that the X-ray luminosity ($L_X$) of a T-Tauri star depends on its mass and age, with an average value of $\log(L_X) = 30.37 \pm 0.07$ ergs s$^{-1}$ for 1 M$_\odot$ (solar mass) and 1 Myr—i.e., ~3 orders of magnitude higher than the X-ray luminosity of the contemporary Sun (Preibisch et al. 2005). For pre–main sequence stars in the age range of 0.1–10 Myr and the solar mass range of 0.1 < M < 2 M$_\odot$, the luminosity decreases slightly with stellar age according to $L_x \propto t^{-1/3}$, whereas the decrease is more rapid over longer timescales (Preibisch & Feigelson 2005). This implies that a solar mass star can be considered active for X-ray emission during the first ~10 Myr of its life.

Given astronomical observations of YSOs, it is difficult to imagine how the young Sun could have escaped such enhanced X-ray activity. The question, however, is to know whether this could have left traces in material that is now in the inner part of the solar system. The X-ray luminosity is the sign of enhanced luminosity at all wavelengths, but it is also accompanied by an enhanced emission of protons, α particles, and $^3$He nuclei to energies of 1 MeV per nucleon and higher. The proton luminosity $L_p$ for particles with energies higher than 10 MeV can be scaled to $L_X$ using observations in present-day solar flares according to $L_p(E \geq E_{10}) \approx 0.09 \times L_X$ (Lee et al. 1998). Interactions of these accelerated particles with ambient gas and grains result in nuclear reactions that can theoretically produce a wealth of nuclei, including short-lived radionuclides. Although extinct radionuclides are the topic of this review, it is important to stress that the potential production of some short-lived radionuclides in the accretion disk is only one of the fundamental consequences of an early active Sun (see review by Chaussidon & Gounelle 2006). Irradiation by ultraviolet light could be of key importance to other processes. Self-shielding of ultraviolet light by CO gas (Clayton 2002) was, for instance, one of the mechanisms considered to explain the presence of non-mass-dependent oxygen isotope variations in solar system objects. Extreme ultraviolet light from the T-Tauri Sun was also suggested to be responsible for hydrodynamic escape of the atmosphere from the young Earth (Dauphas 2003, Hunten et al. 1987, Pepin 1991, Sekiya et al. 1980).

The finding of short-lived $^{10}$Be ($t_{1/2}$ = 1.39 Myr), in the form of $^{10}$B excesses in a CAI corresponding to $^{10}$Be/$^9$Be = 8.8 ± 0.6 × 10$^{-4}$ (McKeegan et al. 2000), is proof that irradiation occurred and left traces in solar system material, as this nuclide cannot be produced by other processes. In the same inclusion, large $^7$Li excesses also were detected, which can be explained by the in situ decay of short-lived $^7$Be ($t_{1/2}$ = 53 days) with $^7$Be/$^9$Be = 6.1 ± 1.6 ×

$10^{-3}$ at the time of crystallization of the CAI (Chaussidon et al. 2006). Because of its extremely short half-life and the complex geochemical behavior of daughter $^{7}$Li, identification of $^{7}$Be in CAIs is not definitive, and further work will be required to ascertain its presence in the early solar system. Li-Be-B elements are not produced in significant amounts by Big Bang nucleosynthesis or stellar nucleosynthesis but instead result from spallation reactions between galactic cosmic-ray protons and O and C nuclei in the interstellar medium. As a consequence, their cosmic abundances are $\sim 10^{6}$--$10^{9}$ times lower than those of neighboring light elements, which makes them sensitive to production through spallation reactions by early solar cosmic rays (Reeves 1994, Reeves et al. 1970). Berryllium-10 is widespread in refractory phases (in hibonites and various types of refractory inclusions) formed early in the solar disk, with $^{10}$Be/$^{9}$Be ratios ranging from $\sim 3 \times 10^{-4}$ to $\sim 1.8 \times 10^{-3}$ (Liu et al. 2009, 2010; MacPherson et al. 2003; Marhas et al. 2002; McKeegan et al. 2000; Sugiura et al. 2001). The former presence of $^{10}$Be has not been demonstrated in chondrules, probably because of low Be/B fractionation, but investigators have found large Li and B isotopic variations that likely result from the addition of spallogenic Li and B produced together with $^{10}$Be (and $^{7}$Be) (Chaussidon & Robert 1995, 1998). Similarly, $^{21}$Ne cosmogenic excesses attributed to early solar system irradiation processes have long been known in olivines (a ferromagnesian silicate) from different meteorites (Caffee et al. 1987, Hohenberg et al. 1990), and their presence recently has been established in meteoritic chondrules (Das et al. 2010).

In addition to the case for $^{10}$Be, the case for an irradiation origin of short-lived $^{36}$Cl ($t_{1/2}$ = 0.3 Myr) is also strong. This is because the initial $^{36}$Cl/$^{35}$Cl ratio inferred to explain the $^{36}$Cl excesses found in late-stage alteration products of CAIs exceeds $\sim 10^{-4}$, which is too high to be accounted for by any stellar source (Hsu et al. 2006, Jacobsen et al. 2009, Lin et al. 2005). The situation is less clear for $^{26}$Al, $^{41}$Ca, and $^{53}$Mn, although an irradiation origin of $^{26}$Al was considered a possibility when it was discovered (Heymann & Dziczkaniec 1976; Lee et al. 1976, 1977). Different calculations have been performed assuming various irradiation scenarios, either at asteroidal distances assuming no shielding by nebular gas (Goswami et al. 2001, Marhas et al. 2002) or at distances close to the Sun (Gounelle et al. 2001, 2006; Leya et al. 2003) in the framework of the X-wind model. In this model, proto-CAIs are irradiated in the reconnection ring at the inner edge of the accretion disk before being launched to asteroidal distances by a magnetocentrifugal outflow known as the the X-wind (Shu et al. 1996, 2001). Although these calculations differ in their results because of the uncertainty in the parameters considered (e.g., fluence, composition of the target), they agree on the general picture that whereas $^{36}$Cl, $^{53}$Mn, and $^{41}$Ca can be produced together with $^{10}$Be at their meteoritic abundances, $^{26}$Al is generally underproduced by a factor of 2--5 unless specific conditions are assumed (e.g., $^{3}$He-rich impulsive flares). This factor holds if no limitation is considered in the total energy available for irradiation or in the total mass of irradiated material. In consideration of such limitations, the conclusion is that although irradiation must contribute locally to the budget of early solar system $^{26}$Al, it is probably a minor component in the overall disk inventory . First, $^{26}$Al and $^{10}$Be do not correlate in hibonites and CAIs (Chaussidon & Gounelle 2006, Marhas et al. 2002, Liu et al. 2009). Second, the systematics of the Mg isotopic composition of Earth, chondrules, and CAIs implies that $^{26}$Al was homogeneously distributed (at the $\pm 10$ % level) in the accretion disk at the time of formation of type B (once-molten, coarse-grained) CAIs (Villeneuve et al. 2009). This observation is difficult to reconcile with a local irradiation scenario. Finally, energetic constraints obtained from the observations of YSOs limit the amount of $^{26}$Al that could have been produced by irradiation (Duprat & Tatischeff 2007, 2008). A homogeneous distribution of $^{10}$Be and $^{26}$Al in the accretion disk implies the production of $3 \times 10^{-8}$ $M_\oplus$ (Earth mass) of $^{10}$Be and $7 \times 10^{-4}$ $M_\oplus$ of $^{26}$Al. This is the most conservative estimate made for a disk with a mass of 0.5 $M_\odot$, a dust/gas ratio of $10^{-2}$, a $^{10}$Be/$^9$Be ratio of $10^{-3}$, a $^{26}$Al/$^{27}$Al ratio of $5 \times 10^{-5}$, and Be and Al in solar or chondritic concentrations. If irradiation of solids close to the Sun is assumed (the most favorable case with no shielding of protons), the required amount of $^{10}$Be is obtained for a 10-MeV proton fluence of $4.3 \times 10^{43}$ erg, which corresponds to 3 Myr of irradiation around a young Sun with an X-ray luminosity of $5 \times 10^{23}$ J s$^{-1}$ ($10^3$ times that of the present-day Sun). Under these conditions, the amount of $^{26}$Al produced is ~500 times lower than the required amount (Duprat & Tatischeff 2007, 2008). If irradiation products were not distributed homogeneously in the entire disk but just in 80 $M_\oplus$, i.e., in the portion of the disk corresponding to the rocky component of a minimum-mass solar nebula, the irradiation time required to make the correct amount of $^{10}$Be is reduced from 3 to 0.1 Myr. However, $^{26}$Al is still underproduced by a factor of ~30 if it is to make 80 $M_\oplus$ of chondrites with $^{26}$Al/$^{27}$Al = $5 \times 10^{-5}$.

The presence of $^{10}$Be in the refractory components of chondrites (CAIs, hibonites) anchors the formation of these minerals formed by condensation from the hot (>1500 K) nebula gas to the pre–main sequence evolution of the Sun (the T-Tauri phase shown in **Figure 3**). The variations in $^{10}$Be/$^9$Be observed between some igneous refractory inclusions ($8.8 \pm 0.6 \times 10^{-4}$) and more refractory hibonite grains ($5.3 \pm 1.0 \times 10^{-4}$)—which, taken at face value, would indicate hibonites to be ~1.1 Myr younger than some CAIs—probably do not indicate time differences. Instead, the variations likely reflect heterogeneities in the distribution of $^{10}$Be, in agreement with irradiation processes taking place during short and repetitive flare events (Liu et al. 2009). The presence of irradiation products (cosmogenic Li, B, and $^{21}$Ne) in chondrules that are between ~1.2 and ~4 Myr younger than CAIs (Kita et al. 2005a, Villeneuve et al. 2009) is consistent with the expected duration of the pre–main sequence phase of the Sun.

## 4. TIME ZERO, DUST CONDENSATION AND AGGLOMERATION

Most primitive meteorites contain refractory inclusions several tens of micrometers to a few centimeters in size that have ceramic-like mineralogy and chemistry (Christophe Michel-Lévy 1968, Grossman 1972, MacPherson 2005). Equilibrium thermodynamic calculations show that the chemical compositions of these CAIs are consistent with high-temperature condensation (~1,300 to 1,500 K, depending on the total pressure) from a gas of solar composition (Grossman 1972, Grossman et al. 2000). CAIs must have formed early and close to the Sun (e.g., <2 Myr after formation of the Sun and <<2 AU), when and where the temperatures were sufficiently high for the most refractory elements to be in the gas phase (Boss 1998, Ciesla 2010, D'Alessio et al. 2005, Makalkin & Dorofeeva 2009). These products of inner solar system chemistry can be found in cometary material (Simon et al. 2008), indicating large-scale transport of dust in the protoplanetary disk (Ciesla 2007). The absolute age of refractory inclusions can be estimated using $^{206}$Pb-$^{207}$Pb geochronology (Allègre et al. 1995, Amelin et al. 2002). In almost all studies, $^{206}$Pb-$^{207}$Pb ages were calculated assuming a constant $^{238}$U/$^{235}$U ratio of 137.88, but recent work has shown that this value could vary from one CAI to another, possibly because of decay of short-lived radionuclide $^{247}$Cm ($t_{1/2}$ = 15.6 Myr) (Table 1; see also Brennecka et al. 2010). Bouvier & Wadhwa (2010) did not measure directly the U isotopic composition in the CAI that they analyzed; instead, they estimated it on the basis of the observed chemical composition and obtained an age of $4,568.2^{+0.2}_{-0.4}$ Myr. Amelin et al. (2010) reported a $^{206}$Pb-$^{207}$Pb age of 4,567.18 ± 0.50 Myr for a different CAI in which the $^{238}$U/$^{235}$U ratio was also measured. A 1-Myr time span in CAI U-Pb ages due to reheating episodes in the accretion disk cannot be excluded.

The $^{26}$Al-$^{26}$Mg systematics (**Figure 2**) in CAIs can be used to study the timing of dust condensation and agglomeration (**Figure 4**). Fractionation of the Al/Mg ratio in bulk CAIs was established during condensation, as Mg was more volatile than Al and was not totally condensed. Thus, the initial $^{26}$Al/$^{27}$Al ratio inferred from measurements of bulk CAIs (in CV chondrites, a class of CAI-rich carbonaceous chondrites that are named after the Vigarano meteorite) reflects the time when refractory dust condensed from gas in the inner part of the protoplanetary disk. Bulk CAI measurements define a single fossil isochron with a slope corresponding to an initial $^{26}$Al/$^{27}$Al ratio of 5.23 ± 0.13 × 10$^{-5}$ (Jacobsen et al. 2008, Town 2008). A notable feature is that within uncertainties, all CAIs from CV chondrites define the same isochron, indicating that they all condensed within a time span of less than 20 kyr (Jacobsen et al. 2008, Thrane et al. 2006, Town 2008). MacPherson et al. (2010a) measured an internal $^{26}$Al-$^{26}$Mg isochron in a fine-grained CAI (<10-μm grain size) with a microstructure indicative of direct condensation from nebular gas. In that object, fractionation of the Al/Mg ratio between minerals must have occurred during condensation, and the initial $^{26}$Al/$^{27}$Al ratio of 5.27 ± 0.17 × 10$^{-5}$ must record that time. This is identical to the value defined by bulk CAI measurements, and it confirms the conclusion that condensation of refractory dust in the inner part of the disk occurred over a very short duration. This, together with the fact that they formed early, makes CAIs perfect time anchors for early solar system chronology.

**< PLEASE INSERT FIGURE 4 HERE>**

**Figure 4** Chronology of refractory dust condensation, agglomeration, and melting in the inner solar system. The principles behind the use of the $^{26}$Al-$^{26}$Mg dating method in meteorites are presented in **Figure 2**. During cooling of the protosolar nebula, elements condensed into refractory dust <10 μm in size. This dust was then agglomerated into millimeter- to centimeter-sized objects. Some of these dust agglomerates subsequently became molten and crystallized and thus assumed a rounded shape. Isochrons calculated from measurements of bulk CAIs (Jacobsen et al. 2008) and individual minerals in a fine-grained CAI (MacPherson et al. 2010a) give initial $^{26}$Al/$^{27}$Al ratios of ~5.2 × 10$^{-5}$. This corresponds to the time when Al was fractionated from Mg owing to incomplete condensation of nebular gas into CAIs. An internal isochron calculated from measurements of individual minerals in an igneous (molten) CAI gives an identical initial $^{26}$Al/$^{27}$Al ratio of ~5.2 × 10$^{-5}$ (MacPherson et al. 2010b). This corresponds to the time when the centimeter-sized object was melted last and crystallized. The fact that the initial $^{26}$Al/$^{27}$Al ratios are identical within uncertainties for these two events (same slopes in δ$^{26}$Mg* versus $^{27}$Al/$^{24}$Mg diagrams) indicates that it took less than 50 kyr for micrometer-sized dust to agglomerate into centimeter-sized objects.

Not all CAIs represent pristine nebular condensates. Many were melted after they were formed, possibly by nebular shock waves (Ciesla & Hood 2002, Desch & Connolly 2002, Hood & Horanyi 1993, Richter et al. 2006) or by passage near the Sun and launch to the meteorite-forming region by the X-wind (Shu et al. 1996). This melting episode gives us the opportunity to study the timescale of dust agglomeration. Indeed, the time of melting and subsequent crystallization represents a robust upper limit on the time of agglomeration. To estimate the initial $^{26}$Al/$^{27}$Al ratio at the time of crystallization of igneous (once molten) CAIs, one can calculate an internal isochron on the basis of measurements of individual minerals (Lee et al. 1976, 1977). MacPherson et al. (2010b) measured at high precision the initial $^{26}$Al/$^{27}$Al ratio of four igneous CAIs and obtained values that range from $4.24 \pm 0.36 \times 10^{-5}$ to $5.17 \pm 0.31 \times 10^{-5}$. The highest value corresponds to a time difference between condensation ($^{26}$Al/$^{27}$Al$_0$ = $5.23 \pm 0.13 \times 10^{-5}$; Jacobsen et al. 2008) and melting/crystallization ($^{26}$Al/$^{27}$Al$_0$ = $5.17 \pm 0.31 \times 10^{-5}$; MacPherson et al. 2010b) of less than 50 kyr. Some dust agglomerates could have formed later, but the study of CAIs shows that it took less than 50 kyr for condensates less than ~10 μm in size to agglomerate into centimeter-sized objects.

As CAIs represent ideal time anchors for early solar system chronology, one would like to put their formation into the broader context of solar system birth. In the inner part of the disk (within ~1 AU), the temperature profile is almost isothermal. This is due to the thermostatic effect of ferromagnesian silicate dust vaporization, which modulates the opacity (Boss 1998, Morfill 1988). Sufficiently high temperatures for the precursors of CAIs to condense could have been sustained in that region of the disk for several million years (Ciesla 2010, D'Alessio et al. 2005, Makalkin & Dorofeeva 2009). This duration is much longer than the time span of <20 kyr for condensation of the precursors of CAIs inferred from $^{26}$Al-$^{26}$Mg systematics (Jacobsen et al. 2008, MacPherson et al. 2010a, Thrane et al. 2006, Town 2008). Ciesla (2010) studied the dynamic fate of CAIs formed at different times in the disk and concluded that early-formed inclusions should dominate the inventory of CAIs found in primitive chondrites. The reasons are that (*a*) most inclusions are formed early, before the disk is drained of its mass and (*b*) early-formed inclusions are dispersed in the disk more easily by outward transport associated with angular momentum redistribution. Therefore, the time that refractory inclusions condensed marks the time when the protoplanetary disk reached its maximum size and when temperatures in the inner part first reached ~1,300 to 1,500 K. This corresponds to the transition between class I and class II YSOs (examples of midplane temperature profiles in class I and II sources can be found in D'Alessio et al. 2005 and Whitney et al. 2003). This view is consistent with the observation that CAIs contain $^{10}$Be, which must have been produced by irradiation when the Sun was a T-Tauri star (see Section 3).

The timescale of grain coagulation in the inner part of the solar system estimated from the study of CAIs (<50 kyr to grow from <10 μm to ~5 mm) can be compared with model predictions and spectroscopic observations of grain-size distributions in YSOs. Dust grains in the interstellar medium have size distributions that peak at 0.2--0.3 μm, with most grains having sizes of less than a micrometer (Mathis et al. 1977, Weingartner & Draine 2001). The theory of dust coagulation in protoplanetary disks predicts that those grains should grow rapidly (Dullemond & Dominik 2005, Weidenschilling 1980). For example, if one were to incorporate the physics of coagulation by Brownian motion, differential settlings to the midplane, and turbulent mixing, the expectation is that the grains should grow from <0.5 μm to 1 cm on a timescale of 1,000 years (Dullemond & Dominik 2005). Models also predict a rapid depletion of the smallest grains, which is not seen in T-Tauri disks, indicating that small grains must be replenished constantly by fragmentation of larger dust agglomerates.
Astronomical observations of YSOs rely on two regions of the spectrum to examine grain growth in protoplanetary disks. The mid-infrared region (especially the Si-O stretching mode feature near 10 μm) can probe grain growth up to several micrometers at the surface of the disk within a few astronomical units of the central star (Bouwman et al. 2001, Kessler-Silacci et al. 2006, Przygodda et al. 2003). Infrared observations of pre–main sequence stars less than 1.5 Myr old show that dust particles grow rapidly to micrometer-sized agglomerates (Kessler-Silacci et al. 2006). The (sub)millimeter- to centimeter-wavelength region can probe grain growth up to centimeter-sized grains in the cold midplane of the outer disk (Calvet et al. 2002, Lommen et al. 2010, Ricci et al. 2010, Rodmann et al. 2006). Again, observations indicate rapid growth to millimeter-/centimeter-sized grains. Ricci et al. (2010) argued that this occurs before a YSO enters the class II evolutionary stage. Kwon et al. (2009) found evidence for grain growth in even younger, class 0, protostellar systems. Thus, astronomical observations are in good agreement with the conclusions based on dating of meteorite refractory inclusions and their constituents.

## 5. FROM DUST TO PLANETESIMALS

Planetesimals can be defined as objects formed in the accretion disk that were large enough (kilometer sized and up) to escape gas drag. A common view is that the production of planetesimals 1–10 km in size is the first step toward the formation of terrestrial planets (see Section 6). In this scenario, growth beyond ~1 m in size, the so-called meter-size barrier, has always been difficult to explain. Whereas models of dust-gas interactions predict that growth to meter-sized objects can proceed rapidly (i.e., in 100--1,000 years at 1 AU and $6$--$7 \times 10^4$ years at 30 AU; Dullemond & Dominik 2005, Weidenschilling 1997) by collisions between smaller-sized aggregates (see Dominik et al. 2007 and references therein), further growth is problematic. Indeed, fragmentation and bouncing tend to prevail over growth in that size range (Zsom et al. 2010). This problem is exacerbated by the fact that meter-sized bodies are expected to have a short dynamic lifetime. Objects orbiting the Sun are bathed in nebular gas that rotates more slowly than the Keplerian velocity. As a consequence, they encounter gas drag, which causes an inward drift of the solids. This drift is minimum for small (0.1-μm-sized) particles that are tightly coupled to the gas and for large (kilometer-sized) bodies that move in Keplerian orbits. The drift velocity is predicted to be maximum for meter-sized objects (up to ~10 m s$^{-1}$ radial drift for a nonturbulent nebula and objects of unit density orbiting the Sun between 1 and 30 AU; Cuzzi & Weidenschilling 2006). Thus, inward drift of particles in the meter-size range might occur so quickly as to outpace agglomeration. Turbulence can increase the residence time of the particles but might also counteract their settling onto the midplane (Dubrulle et al. 1994, Weidenschilling 1980).
Different paths to the formation of planetesimals have been proposed recently. All these models use turbulence to promote the formation of clumps of particles that can be dense enough to collapse into planetesimals by gravitational instability. They break the paradigm of a hierarchical growth because they predict direct formation of 100--1,000-km-sized planetesimals. A first class of models (Johansen et al. 2007) supposes that pressure maxima in a turbulent disk can concentrate meter-sized particles, leading to regions of high particle/gas ratios in which the gas is dragged to near-Keplerian speeds by the particles. These regions concentrate particles drifting from further out in the disk, in a process known as streaming instability, and gravitationally bound clumps develop. Calculations show that this process can produce large (~1,000-km-sized) planetesimals in an extremely short time (a few to ~10 orbital periods). A second class of models (Chambers 2010, Cuzzi et al. 2008) considers millimeter-sized particles (i.e., of chondrule size) that can be concentrated in low-vorticity regions of a turbulent disk. In some eddies, the particles reach a high enough concentration to form gravitationally bound clumps that can resist rotational breakup and the ram pressure of the surrounding gas. These clumps can ultimately undergo gradual contraction to form 10--100-km-sized planetesimals. This process, to be efficient, requires dust/gas ratios one to two orders of magnitude higher than the canonical solar ratio. Whereas the rate of formation of planetesimals can be rapid at 10 AU (a few tens of thousands of years), 100-km-sized planetesimals would require a few million years to form in regions inside the snow line (the heliocentric distance beyond which water ice is stable). Independent support for these scenarios comes from simulations of the growth of planetary embryos: these simulations can best reproduce the original size distribution of objects in the asteroid belt if the initial planetesimals have sizes in the range of 100--1,000 km (Morbidelli et al. 2009). All these models constitute a breakthrough in our understanding of the formation of planetesimals and make predictions concerning the size, the age, and the nature of planetesimals, all of which can be tested using meteorites. Abundances of short-lived radionuclides in meteorites demonstrate that primary accretion—by primary, we mean formed along the aforementioned principles of collapse of a gravitationally bound clump of particles—took place very early in the inner solar system, occurred quickly, and lasted at least ~4 Myr after CAIs, until the nebula gas was dissipated. No data exist to date accretion of objects in the outer solar system.

The first constraint on the formation time of planetesimals is given by the $^{182}$W/$^{183}$W ratio of magmatic iron meteorites (see **Figure 5** for an explanation of $^{182}$Hf-$^{182}$W systematics). After correction for cosmogenic effects resulting from exposure to galactic cosmic rays, the $^{182}$W/$^{183}$W ratios of magmatic iron meteorites are equally radiogenic or only slightly more radiogenic than the solar system initial ratio defined by CAIs (Burkhardt et al. 2008; Kleine et al. 2005, 2009; Markowski et al. 2006, Qin et al. 2008). This implies that the parent bodies of magmatic iron meteorites (bodies of ~10--200 km in size; Chabot & Haack 2006) accreted ~0--1.5 Myr after the formation of CAIs (although analytical errors permit accretion to begin a few hundred thousand years earlier than the formation of CAIs). Any body exceeding ~1 km formed within one million years of the formation of CAIs would likely undergo rapid metal-silicate differentiation in response to the heat released by the decay of $^{26}$Al. For instance, a 5-km-diameter planetesimal (with $^{26}$Al/$^{27}$Al = 5 × 10$^{-5}$ and initial temperature $T$ at 100 K) would reach 1,500 K in its core within ~4 × 10$^5$ years, whereas its surface would remain at $T < 400$ K (e.g., Ghosh et al. 2006, Hevey & Sanders 2006, LaTourette & Wasserburg 1998). Basaltic angrites, which, according to their $^{26}$Al, $^{182}$Hf, and Pb-Pb systematics (Amelin 2008, Baker et al. 2005, Markowski et al. 2007, Schiller et al. 2010, Spivak-Birndorf et al. 2009), are the oldest non-iron-differentiated meteorites (crystallization ages of ~5 Myr after CAI formation) and are likely fragments of planetesimal crusts formed within ~2--3 Myr of CAI formation. Such a rapid accretion of bodies ~10--100 km in size shows that the mechanisms described by Johansen et al. (2007) and Cuzzi et al. (2008) could have operated in the accretion disk. Early formation of large planetary objects might have induced collisional evolution and scattering of the rest of the population of planetesimals (Bottke et al. 2006).

< INSERT FIGURE 5 HERE>

Figure 5 Principles behind the use of $^{182}$Hf-$^{182}$W systematics ($t_{1/2}$ = 8.9 Myr) for dating core-mantle segregation in the context of a simple two-stage model of core formation. In a first stage ($t < t_c$, where $t_c$ stands for time of core formation), the planetary body is assumed to follow a chondritic evolution. In chondrites, metal (*yellow*) is intimately mixed with silicate (*gray*) at the subcentimeter scale. In a second stage ($t = t_c$), the metal melts and is segregated from the silicate mantle to form a core. Given that Hf is lithophile whereas W is moderately siderophile, the Hf/W ratio of the mantle is higher than that of the core or chondrites. After core formation, $^{182}$Hf continues to decay into $^{182}$W, resulting in a more radiogenic (higher) $^{182}$W/$^{183}$W ratio in the mantle compared with that of the chondrites. The Hf/W ratio of the core is ~0, so its W isotopic composition is identical to the chondritic composition at $t_c$. Measuring the $^{182}$W/$^{183}$W ratio of planetary mantles ($\epsilon^{182}$W$_{mantle}$ in the following equation) and knowing their Hf/W ratios ($f_{mantle}^{Hf/W}$), one can determine the time that the composition of the mantle departed from the chondritic evolution, which corresponds to the time of core formation

$$(t_c), \epsilon^{182}W_{mantle} = q_W f_{mantle}^{Hf/W} \left(\frac{^{182}Hf}{^{180}Hf}\right)_0 e^{-\lambda t_c}, \epsilon^{182}W_{mantle} = q_W f_{mantle}^{Hf/W} \left(^{182}Hf/^{180}Hf\right)_0 e^{-\lambda t_c}, \text{where}$$

$\epsilon^{182}W_{mantle} = [(^{182}W/^{183}W)_{mantle}/(^{182}W/^{183}W)_{CHUR} - 1] \times 10^4$, $q_W = 10^4 \times (^{180}Hf/^{182}W)_{CHUR}$, and $f_{mantle}^{Hf/W} = (Hf/W)_{mantle}/(Hf/W)_{CHUR} - 1$. The values for the chondritic uniform reservoir (CHUR) represent the bulk composition of Mars (Lodders & Fegley 1997). We have $\epsilon^{182}W_{mantle}$ = +2.57 (Foley et al. 2005, Kleine et al. 2007), $q_W$ = 1.07 × 10$^4$ (Kleine et al. 2007), and $f_{mantle}^{Hf/W}$ = 3.38 (Dauphas & Pourmand 2011). The initial ($^{182}$Hf/$^{180}$Hf)$_0$ ratio of 9.72 ± 0.44 × 10$^{-5}$ is well established from measurements of refractory inclusions (Burkhardt et al. 2008), and the decay constant $\lambda_{^{182}Hf}$ is 0.078 Myr$^{-1}$ (Vockenhuber et al. 2004). Applying this simple model to the martian mantle would give a time of core formation of ~4.0 Myr. However, this model may be inappropriate if core formation in embryos and planets was a protracted phenomenon that tracked the accretion history.

The second constraint on the formation time of planetesimals comes from chondrules, which show that chondrite parent bodies accreted over a longer time period of 3–4 Myr, as predicted in the dynamic models of Cuzzi et al. (2008) and Chambers (2010). Pb-Pb ages, short-lived $^{26}$Al, and $^{53}$Mn indicate that chondrules crystallized ~1--3 Myr after the formation of CAIs (e.g., Kita et al. 2005a and references therein). High-precision $^{26}$Al isochrons (see **Figure 3c**) in type II chondrules from Semarkona show that they have crystallization ages from 1.2 to 4 Myr (Villeneuve et al. 2009). All the existing internal $^{26}$Al isochrons in chondrules taken at face value (**Figure 6**), whatever their precisions are, show that chondrites did not form until ~4 Myr after CAI formation. Detailed geochronological studies of some chondrites (H-type chondrites) have shown that they probably came from a parent body that was ~100 km in diameter (e.g., Trieloff et al. 2003). The samples that come from near the center of that parent body have the same chemical compositions and same constituents as those that come from more external regions. As discussed by Cuzzi et al. (2008), this is consistent with the idea that such planetesimals were formed by collapse of a high-density

(Cuzzi & Alexander 2006) clump of gravitationally bound chondrules, refractory inclusions, and fine dust. Chondrites do not look like rubble piles of unrelated meter-sized boulders. Thus, primary planetesimals could have formed throughout the accretion history of the disk.

**< PLEASE INSERT FIGURE 6 HERE>**

---

Figure 6 Constraints from short-lived radionuclides on the timing of primary accretion. Relative ages are derived from $^{182}$Hf for magmatic iron meteorites (Burkhardt et al. 2008; Kleine et al. 2005, 2009; Markowski et al. 2006, 2007; Qin et al. 2008), $^{26}$Al for chondrules [total of 112 chondrules, including 68 for unequilibrated ordinary chondrites (*solid curve*) and 44 from carbonaceous chondrites (*dashed curve*); data from Hsu et al. 2003; Hutcheon & Hutchison 1989; Hutcheon et al. 2000; Kita et al. 2000, 2005b; Kunihiro et al. 2004; Kurahashi et al. 2008; Mostéfaoui et al. 2000; Nagashima et al. 2007, 2008; Rudraswami & Goswami 2007; Rudraswami et al. 2008; Russell et al. 1996; Sugiura & Krot 2007; Yurimoto & Wasson 2002], and $^{26}$Al for eucrites (originating most probably from the asteroid Vesta) and angrites (Baker et al. 2005, Bizzarro et al. 2005, Schiller et al. 2010, Spivak-Birndorf et al. 2009). The boxes shown for magmatic irons, angrites, and eucrites correspond to the duration from accretion to global melting ($^{26}$Al and $^{182}$Hf give access to differentiation ages for these bodies, and the accretion age is back-calculated from modeling of heating by $^{26}$Al; e.g., Qin et al. 2008). Contrary to the classical view of planetary formation, it is clear that accretion started very early with some parent bodies of magmatic iron meteorites that could be as old as calcium-, aluminum-rich inclusions (CAIs), and that it continued most probably for up to 4 Myr. The age distribution of chondrules could be interpreted either as the reflection of some major episodes of formation between ~1 and 4 Myr after CAI formation or as the result of a biased sampling (or preservation) of the population existing in the disk at the time of chondrite formation. Whatever the exact origin of chondrules' age distribution, this demonstrates the existence of processes able to preserve chondrules from accreting to the Sun for several million years. Chondrites were formed late (most likely ~4 Myr after CAI). Secondary planetesimals can be made at any time from collisions between primary planetesimals. CB chondrites (a class of metal-rich carbonaceous chondrites that are named after the Bencubbin meteorite; see Section 5) are probably an example of such an episode that happened late, ~5.5 Myr after CAI formation (Krot et al. 2005a).

---

Secondary planetesimals can be defined as those formed by collisions between primary planetesimals. Different hints for very early and late collisions and fragmentation of planetesimals exist. The first hint comes from Mg-rich olivines in a peculiar type of chondrule (the so-called type I chondrule). These olivines are present in chondrules as relict clasts. They have a granoblastic texture (a rock texture in which equant crystals meet at 120° triple junctions) analogous to that of terrestrial mantle rocks (Libourel & Krot 2007). This suggests that they could be fragments of the mantles of disrupted planetesimals that accreted very early, akin to the parent bodies of magmatic iron meteorites or angrites. Oxygen isotope systematics supports an origin from several planetesimals caracterized by different oxygen isotopic compositions (Libourel & Chaussidon 2011). The unradiogenic Mg isotopic composition of several of these olivines is consistent with an early formation, within ~0.1--1 Myr of CAI formation (Villeneuve et al. 2011). The second hint comes from CB chondrites, a class of metal-rich carbonaceous chondrites that are named after the Bencubbin meteorite. The CB chondrites contain chondrules that have young Pb-Pb ages (~5.5 Myr after CAIs from CV chondrites; Krot et al. 2005a) and CAIs that show mass fractionation of Mg isotopes along with a low abundance of $^{26}$Al ($^{26}$Al/$^{27}$Al < 4.6 × 10$^{-6}$; Gounelle et al. 2007). This is in agreement with formation in a plume generated by an impact between planetesimals or early-formed embryos.

All these observations are in agreement with the latest dynamic models and indicate that planetesimals can form over different timescales (the ones formed early and rapidly were molten and differentiated, whereas those formed later from millimeter-sized dust remained undifferentiated). The view that the evolution of the nebula occurred in discrete chronological eras (i.e., agglomeration of micrometer-sized condensates, formation of meter-sized objects, and formation of kilometer-sized planetesimals) appears incorrect, as planetesimal formation spanned several millions of years and overlapped with embryo formation.

## 6. FROM PLANETESIMALS TO MARS, A POSSIBLE PLANETARY EMBRYO

In the classical model of planetary formation, the growth of planetesimals (from ~1 to possibly 1,000 km in diameter) into planetary embryos (1,000–5,000 km) proceeds in two stages. As planetesimals collide with one another, the largest bodies can exert a stronger gravitational focus on the surrounding objects, and their growth can outpace that of the smaller planetesimals (dM/dt ∝ M$^{4/3}$) in a process known as runaway growth (Greenberg et al. 1978, Wetherill & Stewart 1993). However, as they grow in size, the embryos dynamically excite the swarm of planetesimals around them. Planetary bodies involved in close encounters are less likely to collide when the relative velocities increase, so the largest bodies continue to grow, but they grow at a slower pace in a process known as oligarchic growth (Ida & Makino 1993, Kokubo & Ida 1998). As discussed in Section 5, this simple picture has been questioned by two recent studies that proposed new models for the formation of planetesimals [see the review by Chiang & Youdin (2010)]. Johansen et al. (2007) showed that in turbulent circumstellar disks, dust could concentrate in high-pressure regions (Haghighipour & Boss 2003) for periods of time long enough to allow gravitationally bound clumps of meter-sized objects to collapse into planetesimals. Cuzzi et al. (2008) and Chambers (2010) showed that millimeter-sized particles such as chondrules could concentrate in the cascade of eddies associated with turbulence in the disk. Again, simulations showed that such clumps could survive for a period of time long enough to allow particles to sediment into planetesimals. A common feature of both models is that the planetesimals may have started larger (i.e., ~100–1,000 km; Chambers 2010, Cuzzi et al. 2008, Johansen et al. 2007) than previously thought (i.e., ~1 km; Goldreich & Ward 1973), thus largely bypassing the stage of runaway growth. These results have important implications for the mode and timescale of embryo formation. Using a model that started with large planetesimals (100–1,000 km) created over a 2-Myr interval, Morbidelli et al. (2009) showed that the formation of embryos could take place on a timescale of a few million years (i.e., within ~3 Myr).

Establishing the chronology of embryo formation using extinct radionuclides is not straightforward because most of this material was incorporated into the terrestrial planets during their accretion. However, dynamic simulations of terrestrial planet formation have difficulties explaining the small mass of Mars compared with the masses of Earth and Venus (Chambers & Cassen 2002, Raymond et al. 2009, Wetherill 1991). One possibility is that Mars is actually a planetary embryo that escaped ejection and collision with other embryos

(Chambers 2004, Chambers & Wetherill 1998). If that is correct, its accretion timescale should give us some clues about the formation time of embryos in the inner part of the solar system. The best tool to investigate this question is the radionuclide $^{182}$Hf, which decays to $^{182}$W with a $t_{1/2}$ of 8.9 Myr (**Figure 5**; see also Dauphas & Pourmand 2011; Foley et al. 2005; Jacobsen 2005; Kleine et al. 2007, 2009; Lee & Halliday 1997; Nimmo & Kleine 2007). Hf is a lithophile (silicate rock–loving) element, meaning that when planetary objects melted and metallic cores formed, it was retained in the silicate mantles. Conversely, W is a moderately siderophile (metal-loving) element, meaning that it was largely scavenged into planetary cores during differentiation. As a result of these contrasting behaviors, the mantles of differentiated objects have high Hf/W ratios relative to bulk planets represented by chondrites, whereas their cores have Hf/W ~ 0. If planetary bodies differentiated early, while $^{182}$Hf was still alive (within ~50 Myr of the formation of the solar system), their mantles should show excess radiogenic $^{182}$W relative to that in chondrites owing to the decay of $^{182}$Hf. In contrast, if core-mantle segregation occurred after $^{182}$Hf had decayed, the mantles should have $^{182}$W isotopic abundance identical to that in chondrites. Thus, by measuring the $^{182}$W/$^{183}$W and Hf/W ratios of the martian mantle, one can estimate a model age of core formation, which is related to the timescale of accretion. Because the W isotopic variations are tiny, one uses the ε notation to report them:

$$\varepsilon^{182}W_{mantle} = [(^{182}W/^{183}W)_{mantle} / (^{182}W/^{183}W)_{CHUR} - 1] \times 10^4.$$

(CHUR represents the composition of chondrites, which are the assumed building blocks of Mars.) Measurements of the Shergottite-Nakhlite-Chassignite group of meteorites, which are thought to come from Mars (Bogard & Johnson 1983 and references therein), give the W isotopic composition of the martian mantle relative to chondrites as $\varepsilon^{182}W = +2.57$ (Foley et al. 2005, Kleine et al. 2007). The other important ingredient for applying the $^{182}$Hf-$^{182}$W system to Mars is the Hf/W ratio of its mantle, which can be expressed using a fractionation factor,

$$f_{mantle}^{Hf/W} = (Hf/W)_{mantle}/(Hf/W)_{CHUR} - 1.$$

Again, CHUR represents chondritic composition, assumed to represent the composition of bulk Mars. Nimmo & Kleine (2007) concluded that the calculated age of core formation was uncertain owing to a large uncertainty in the value of $f_{mantle}^{Hf/W}$. However, Dauphas & Pourmand (2011) have redetermined this parameter using a new method, and they have obtained a precise value of $f_{mantle}^{Hf/W} = 3.38$, which provides new insights into the accretion history of Mars.

There is considerable uncertainty about the mode of core formation in embryos and planets (Ricard et al. 2009, Rubie et al. 2003). The simplest scenario is that core formation occurred instantaneously after the embryo or planet completed its accretion (**Figure 5**). For Mars, this gives a time of core formation of 4.0 Myr after the birth of the solar system (see equation 14 of Jacobsen 2005), but this model may not be realistic. Instead, core formation may have been a protracted phenomenon that tracked the accretion of the planet. This alternative model assumes that each time a planetesimal is accreted, its metal and silicate portions are fully equilibrated with the mantle of the growing embryo. Thus, the increment of metal that is subsequently removed to the core has the same W isotopic composition as that of the martian mantle at that time (Halliday et al. 1996, Jacobsen 2005, Jacobsen & Harper 1996). In this scenario, the core grows in constant proportion with the mass of Mars and therefore tracks the accretion of the planet. The assumption of complete equilibration between accreted planetesimals and the mantle of the growing embryo (Mars) is reasonable given the difference in scale between the impactor and the target and the fact that the target may have been molten (Nimmo & Agnor 2006). The accretion history is parameterized using the following equation (Jacobsen & Harper 1996, Wetherill 1986),

$$M = M_{final}\left(1 - e^{-t/\tau}\right),$$

where $M_{final}$ is the present mass of Mars and $\tau$ is the accretion timescale. Given $\tau$ and knowing $f_{mantle}^{Hf/W}$, calculating $\varepsilon^{182}W_{mantle}$ is possible (**Figure 7**). If the accretion timescale was significantly shorter than 2 Myr, one would expect the martian mantle to have more radiogenic $^{182}W$ (higher $\varepsilon^{182}W$) than is measured. Conversely, if the accretion timescale of Mars was much longer than ~2 Myr, one would expect it to have less radiogenic $^{182}W$ (lower $\varepsilon^{182}W$) than is measured. The growth curve is therefore constrained to $\tau \sim 2$ Myr (**Figure 7**; see also Dauphas & Pourmand 2011). Such a short timescale is consistent with the idea that Mars is a stranded planetary embryo and with results of dynamic simulations that indicate formation of planetary embryos within the first few million years of the birth of the solar system (Chambers 2004, Morbidelli et al. 2009). The chronology of planetesimal formation and embryo formation inferred from meteorites (see Sections 5 and 6) indicates that the two processes largely overlapped. Some primary planetesimals such as the parent bodies of H-type chondrites (100 km) were formed a few million years after the formation of CAIs, when Mars (~6,800 km) had already reached a significant size. Embryo formation took place when gas was still present. This is consistent with the core accretion model of giant planet formation that requires accretion of gas onto rocky cores that are formed before dissipation of the nebula (e.g., Mizuno 1980, Pollack et al. 1996).

< PLEASE INSERT FIGURE 7 HERE>

**Figure 7** Accretion history of Mars as derived from $^{182}Hf$-$^{182}W$ systematics of martian meteorites. The growth of Mars is assumed to follow the parameterization $M = M_{final}(1 - e^{-t/\tau})$, where $\tau$ is the accretion timescale (corresponding to the time after the onset of accretion when the planet reached approximately two-thirds of its present size). Given that parameterization, it is possible to estimate the shift in the isotopic abundance of $^{182}W$ in the martian mantle for different values of $\tau$ (equation 65 of Jacobsen 2005), [**AU: Can you find a place in the main text to include this equation? Annual Reviews does not include centered equations within figure captions.

$$\varepsilon^{182}W_{mantle} = q_W \left(\frac{^{182}Hf}{^{180}Hf}\right)_0 f_{mantle}^{Hf/W} \lambda_{182Hf} \int_0^t \left(\frac{1-e^{-\xi/\tau}}{1-e^{-t/\tau}}\right)^{1+f_{mantle}^{Hf/W}} e^{-\lambda\xi} d\xi.$$

$\varepsilon^{182}W_{mantle} = q_W\left(^{182}Hf/^{180}Hf\right)_0 f_{mantle}^{Hf/W} \lambda_{182Hf} \int_0^t \left[\left(1-e^{-\xi/\tau}\right)/\left(1-e^{-t/\tau}\right)\right]^{1+f_{mantle}^{Hf/W}} e^{-\lambda\xi} d\xi.$ The notations and their numerical values for Mars are given in the caption of **Figure 5**. The left panel shows that the measured $^{182}W$ excess in the martian mantle can be explained only by an accretion timescale of τ ~ 2 Myr, which corresponds to the accretion history shown in the right panel (Dauphas & Pourmand 2011).

## 7. FROM EMBRYOS TO EARTH, A TERRESTRIAL PLANET

The outcome of oligarchic growth is the formation of numerous Moon- to Mars-sized embryos surrounded by a swarm of planetesimals (see Section 6). Kokubo & Ida (2000) showed that embryos form at regular orbital distances and that the mass of each embryo depends on the size of its feeding zone and on the surface density of the disk at that location. Terrestrial planets grew by collisions between these planetary embryos; this process is known as chaotic growth because it is entirely stochastic. The transition from oligarchic growth to chaotic growth occurred when the mass locked up in embryos was comparable with that in planetesimals (Kenyon & Bromley 2006). This stage can be followed numerically by simulations in which the trajectories of all embryos are computed (i.e., N-body simulations; Chambers 2001, Chambers & Wetherill 1998, Cox & Lewis 1980, Wetherill 1980). The models are initialized by placing embryos at regular heliocentric distances with masses of individual embryos that depend on the assumed density of the disk with distance and then following the collision merging of these embryos subjected to the laws of gravitation (e.g., figure 1 of Chambers & Wetherill 1998). By convention, in simulations, the larger of two bodies involved in a collision leading to an Earth-like planet is termed the proto-Earth. Early models that tracked only embryos, without planetesimals, found that the planets formed were on orbits that were too eccentric and inclined compared with the real ones. The most likely solution to this problem is that planetesimals were also present when accretion of the terrestrial planets started (Section 6; see also Kenyon & Bromley 2006). In simulations, this reduces the eccentricities of the embryos by dynamic friction and produces terrestrial planets that are more akin to the real ones (Chambers 2001, O'Brien et al. 2006). The presence of the gas giants Jupiter and Saturn— which must have formed within a few million years of the formation of the solar system, before dissipation of nebular gas—also has a strong influence on the outcome of the simulations (Levison & Agnor 2003). State-of-the-art accretion models for Earth predict that the accretion should extend over several tens of million years (**Figure 8**).

< PLEASE INSERT FIGURE 8 HERE>

Figure 8 Predicted $\varepsilon^{182}W$ evolution of the silicate mantle of the proto-Earth (*bottom*) for an N-body simulation of Earth's accretion (*top*) (Kleine et al. 2009, Raymond et al. 2006). This model reproduces approximately the $\varepsilon^{182}W$ value of the silicate Earth (~+2; Dauphas et al. 2002a, Kleine et al. 2002, Schoenberg et al. 2002, Yin et al. 2002). Here, it is assumed that each time an embryo is accreted by the proto-Earth, it is equilibrated with its mantle before metal segregation. A better match between the modeled and measured W isotopic compositions would be obtained by allowing for incomplete equilibration (Allègre et al. 2008, Halliday 2004, Kleine et al. 2009, Nimmo et al. 2010). This model is consistent with the constraint that the Moon-forming impact must have occurred 30--150 Myr after solar system birth (Bourdon et al. 2008, Touboul et al. 2007, Yin et al. 2010). Because the model considers complete equilibration during impact, the $\varepsilon^{182}W$ is partially reset to chondritic composition ($\varepsilon^{182}W = 0$) at each collision with an embryo, which explains the sawtooth pattern in $\varepsilon^{182}W$.

As discussed in the case of Mars (see Section 6), the accretion timescale of Earth can be inferred by applying the principles of $^{182}$Hf-$^{182}$W systematics (for details, see the recent reviews by Jacobsen 2005 and Kleine et al. 2009). Lee & Halliday (1995) were the first to promote the use of $^{182}$Hf-$^{182}$W systematics to constrain the time of core formation on Earth. They had concluded that the silicate Earth had a $^{182}$W/$^{183}$W ratio identical to that of chondrites, implying that the core had formed late, after complete decay of $^{182}$Hf. However, these measurements were later shown to be erroneous (Kleine et al. 2002, Schoenberg et al. 2002, Yin et al. 2002). In 2002, Dauphas et al. (2002a) reinterpreted some earlier W isotope data of enstatite chondrites (Lee & Halliday 2000) and concluded that Earth's mantle had excess radiogenic $^{182}$W corresponding to $\varepsilon^{182}W_{mantle}$ = +2.13 ± 0.53. That same year, Yin et al. (2002), Kleine et al. (2002), and Schoenberg et al. (2002) remeasured some of the carbonaceous chondrites measured earlier by Lee & Halliday (1995) and found that $\varepsilon^{182}W_{mantle} \approx$ +2. The presence of excess $^{182}$W in the terrestrial mantle relative to what is found in chondrites implies that the core formed early, while $^{182}$Hf was still alive.

To quantify how early Earth's core formed, one must rely on models of accretion and core differentiation. As discussed in Section 6, the simplest model is to assume that core formation occurred instantaneously at some time after the birth of the solar system (**Figure 5**). The numerical values to use for Earth's mantle (e.g., assuming enstatite-chondrite-like composition for the bulk Earth) are $\varepsilon^{182}W_{mantle}$ = +2.1 (Dauphas et al. 2002a), $q_W$ = 1.25 × 10$^4$ (Lee & Halliday 2000), and $f^{Hf/W}_{mantle}$ = 21 (Arevalo & McDonough 2008, Dauphas & Pourmand 2011, Newsom et al. 1996). The calculated instantaneous time of core formation on Earth is ~32 Myr (Dauphas et al. 2002a, Kleine et al. 2002, Schoenberg et al. 2002, Yin et al. 2002). One can also calculate the accretion timescale of Earth for an exponentially decaying rate of accretion (see Section 6), and the result is $\tau$ ~ 10 Myr (Jacobsen 2005, Kleine et al. 2009, Yin et al. 2002), which represents the time that Earth would have reached two-thirds of its present size. These results indicate that accretion of Earth took much longer than that of Mars, which is consistent with the idea that their modes of formation differed. However, these two models are overly simplistic, as accretion of Earth corresponded neither to a single differentiation event nor to continuous accretion of small bodies.

For Earth, one can directly use the output of N-body simulations to predict $\varepsilon^{182}$W in the mantle (Jacobsen 2005, Nimmo & Agnor 2006, Nimmo et al. 2010), which can then be compared with the measured value of ~+2. Among the simulations that make Earth-like planets, those that produce the appropriate amount of radiogenic excess $^{182}$W are retained as likely. In the case of Earth, an important difficulty is that growth involved impacts between large bodies, and it is unclear whether the impactor equilibrated with the mantle completely before segregation of metal into the core (Halliday 2004). In an extreme scenario, the cores of impacting embryos involved in collisions would merge without equilibrating at all with the mantle. If this were true, the W isotopic composition of Earth's mantle would give the age of core formation on embryos or planetesimals rather than the age of core formation on Earth. This is not the case (see Sections 5 and 6), so one can rule out this extreme model. Some models assume that only a fraction of the impactor is equilibrated with the mantle of the proto-Earth (Allègre et al. 2008, Dahl & Stevenson 2010, Halliday 2004, Kleine et al. 2009, Nimmo et al. 2010). As shown by Halliday (2004) and Allègre et al. (2008), doing so can help reconcile the age of core formation inferred from $^{182}$Hf-$^{182}$W with the age of core formation inferred from Pb isotopes (Allègre et al. 1982, 1995; Galer & Goldstein 1996) and the age of Earth's atmosphere obtained using $^{129}$I-$^{129}$Xe systematics (Wetherill 1975, Zhang 1998).

An important constraint on the accretion of Earth arises from estimates of the age of the Moon. The current paradigm of Moon formation is that it was produced by a giant impact between a Mars-sized embryo (Theia) and the proto-Earth (Canup & Asphaug 2001). This event occurred at the end of the accretion of Earth, corresponding to an increment of ~10% of its present mass. The lunar mantle has a Hf/W ratio that is higher than that of Earth's mantle (see Touboul et al. 2007, but see also Yin et al. 2010). Thus, if the Moon formed while $^{182}$Hf was still alive, the lunar rocks should show excess $^{182}$W compared with that of Earth's mantle. Instead, Touboul et al. (2007) found that the W isotopic composition of the lunar mantle was indistinguishable from that of Earth. This finding has two important implications: (*a*) the Moon was equilibrated isotopically with the proto-Earth mantle during impact (e.g., Pahlevan & Stevenson 2007), and (*b*) the Moon formed late, after $^{182}$Hf had decayed. Combining this result with other chronological information on the crystallization of the lunar magma ocean (Carlson & Lugmair 1988, Norman et al. 2003), Touboul et al. (2007) and Bourdon et al. (2008) suggested that the Moon formed 50--150 Myr after the birth of the solar system, implying that Earth continued to grow by collisions with embryos for more than 50 Myr. Some numerical simulations give a protracted accretion history of Earth and are consistent with the record of extinct radionuclides in the Earth-Moon system (**Figure 8**) (Kleine et al. 2009). However, our knowledge of the lunar mantle Hf/W ratio is likely less precise than that assumed by Touboul et al. (2007). If, for example, the Hf/W ratio is identical to that of Earth, it would restrict lunar formation to ~30 Myr after CAI formation (Yin et al. 2010), which would be consistent with $^{176}$Lu/$^{176}$Hf evidence from ancient terrestrial zircons (Harrison 2009).

In circumstellar disks of YSOs, the formation of giant planets may be detectable by the development of annular gaps (e.g., Calvet et al. 2002). Some giant planets have been detected directly in debris disks, some as young as 10 Myr old (e.g., Lagrange et al. 2010). Terrestrial planet formation is also amenable to remote observation through the mid-infrared emission from dust produced by catastrophic collisions between embryos and protoplanets (Lisse et al. 2009, Melis et al. 2010). For example, Melis et al. (2010) were able to detect mid-infrared excess emission from a ~80-Myr-old Sun-like star (HD 15407).

## 8. CONCLUSION

Astronomical observations of nascent stars and protoplanetary disks have made tremendous progress in the past decade. They provide insights into the birth of Sun-like stellar systems that complement the vast knowledge already gained by studying the geochemistry of meteorites and terrestrial planets. In this review, we have tried to present how extinct radionuclides can help us understand how the architecture of the solar system was established (**Figure 9**). They are particularly good tracers because they can measure time, work as dosimeters of early solar activity, and give us clues about the context of solar system formation. Future observations of Earth-like extrasolar planets will tell us if the processes and timescales inferred for the solar system also apply to other stellar systems or if we live in a unique place.

**< INSERT FIGURE 9 HERE>**

**Figure 9** Extinct radionuclide perspective on the formation of the solar system. For cosmochemists, time zero is marked by the condensation of refractory dust in the inner part of the disk, which corresponds to the transition between class I and II stellar sources. Less than 50 kyr after condensation, dust agglomerated into millimeter/centimeter-sized objects. Objects of that size (refractory inclusions and chondrules) were present in the disk for ~4 Myr. Fragmentation and bouncing prevented growth by agglomeration beyond the meter-size barrier. Instead, gravitational instability of large clumps of chondrules and refractory inclusions led to the formation of planetesimals up to several hundreds of kilometers in size. Planetesimal formation spanned several million years. Runaway and oligarchic growth produced Mars-sized embryos on a timescale of a few million years. No primary planetesimal was formed after ~4 Myr after solar system birth, which may correspond to the dissipation of the gas and the transition between a class III stellar source and the debris disk. Earth formed by chaotic growth over several tens of million years. Its growth was punctuated by collision with a Mars-sized impactor that formed the Moon at 30--150 Myr after solar system birth. Whenever collisions occurred, debris was produced. The red dashed line gives the conditions required for melting and differentiation by $^{26}$Al decay. It is calculated by equating the heat conduction timescale to $\Delta t_{conduction} = (L/2)^2 \times \rho \times C_p/k$ ($L$ is the diameter of the object, $\rho$ is the density, $C_p$ is the specific heat capacity, and $k$ is the thermal conductivity) with the timescale for adiabatic melting (from equation 8 of Qin et al. 2008):

$\Delta t_{melting} = -(1/\lambda) \ln(1 - \lambda C_p \Delta T e^{\lambda t_{accretion}}/A)$. ($\lambda$ is the decay constant of $^{26}$Al, $t_{accretion}$ is the accretion time of the object, $\Delta T$ is the temperature rise needed to induce melting, and A is the rate of heat production by $^{26}$Al decay at the birth of the solar system.) Thus, the criterion for $^{26}$Al-induced melting in size $L$ versus accretion time $t_{accretion}$ is $L = 2\sqrt{-(k/\lambda \rho C_p) \ln(1 - \lambda C_p \Delta T e^{\lambda t_{accretion}}/A)}$. All the parameters used in calculation of this curve are from Qin et al. (2008). Objects that are less than a few kilometers (<4 km) or that formed more than a few million years (>2 Myr) after solar system birth would not be able to melt and differentiate by $^{26}$Al decay alone (see also Grimm & McSween 1993). The presence of an insulating regolith would allow planetesimals smaller than 4 km to melt.

**SUMMARY POINTS**

1. The presence of very short-lived nuclides such as $^{26}$Al and $^{60}$Fe suggests that the Sun was probably born in a stellar nursery, near massive stars that seeded the young solar system with newly produced radionuclides.
2. For cosmochemists, the clock started ticking when refractory dusts condensed in the inner part of the disk. This corresponds to the transition between class I and II stellar sources.
3. Study of refractory inclusions indicates that micrometer-sized dust agglomerated into millimeter- to centimeter-sized aggregates in less than 50 kyr. This very short timescale is consistent with dust agglomeration timescales inferred from theory and experiments as well as observations of pre–main sequence stars in the mid-infrared. The presence of $^{10}$Be in these inclusions testifies that these objects formed when the Sun was a T-Tauri star.

4. Millimeter- to centimeter-sized dust was abundant in the disk for the first 4 million years of solar system history. Meteorites were found to contain both refractory inclusions and chondrules that formed several million years afterward. This indicates that particles of different ages coexisted in the disk at any given time.

5. Theory tells us that fragmentation and bouncing should prevent meter-sized objects from growing further. Dating of meteorites shows that planetesimals ~10 to 100 km in size formed during the whole lifetime of the protoplanetary disk. The fact that no primary planetesimal is found later than 4 Myr after solar system formation suggests that this time corresponds to dissipation of the gas, marking the transition between a class III stellar source and the debris disk.

6. The planetesimals that formed early (within ~2 Myr of solar system formation) and were sufficiently big (more than ~4 km) were differentiated by the energy released by $^{26}$Al decay. Those formed more than ~2 Myr after solar system formation did not incorporate enough $^{26}$Al to induce melting, and those below ~4 km lost their heat by conduction and were never differentiated (unless they were covered by an insulating regolith).

7. Planetesimals grew by runaway and oligarchic growth into planetary embryos. Mars could be such a stranded embryo, and its accretion timescale of ~2 Myr is entirely consistent with this idea.

8. Earth's accretion had to proceed by chaotic growth and took several tens of million years to complete. It was punctuated by the Moon-forming impact between a Mars-sized embryo and the proto-Earth at 30–150 Myr after solar system birth, which is broadly consistent with the observation of collisional dust in debris disks as old as ~80 Myr.

**ACKNOWLEDGEMENTS**

We thank Eric Quirico, Fred Ciesla, Alessandro Morbidelli, and Reika Yokochi for discussions. Comments from Mark Harrison, Kevin McKeegan, Ming Chang Liu, Ritesh Mishra, Paul Craddock, Thomas Ireland, and undergraduate students from the University of Chicago (Cosmochronology, GEOS 22100) helped improve the quality of the manuscript and were greatly appreciated. This work was supported by a fellowship from the David and Lucile Packard Foundation, the France Chicago Center at the University of Chicago, the National Aeronautics and Space Administration, and the National Science Foundation through grants NNX09AG59G and EAR-0820807 to N.D., and by L'Agence Nationale de la Recherche (ANR grant 08-BLAN-0260-02-T-TauriChem) and the European Research Council under the European Community's Seventh Framework Programme (ERC grant FP7/2007-2013 Grant Agreement #226846) to M.C.

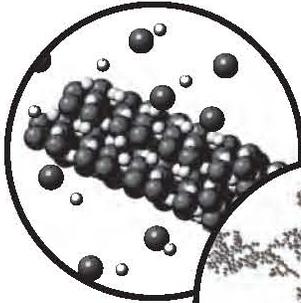
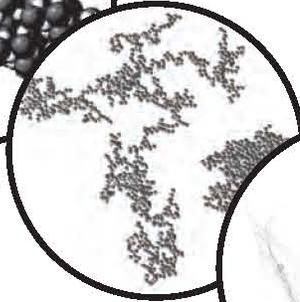
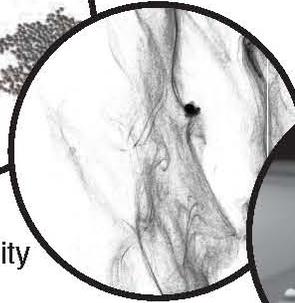
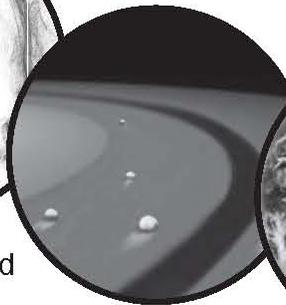
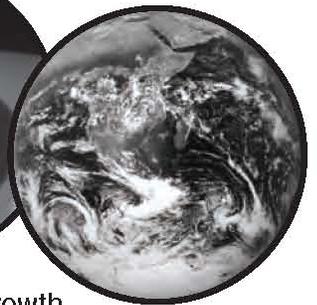

Dust (0.1-10 μm)

Gas condensation, inheritance from ISM

Dust agglomerate (<1 m)

Fractal agglomeration, compaction

Planetesimal (1-1,000 km)

Collisional growth, gravitational instability

Embryo (1,000-5,000 km)

Orderly, runaway, and oligarchic growths

Planet (10,000 km)

Chaotic growth

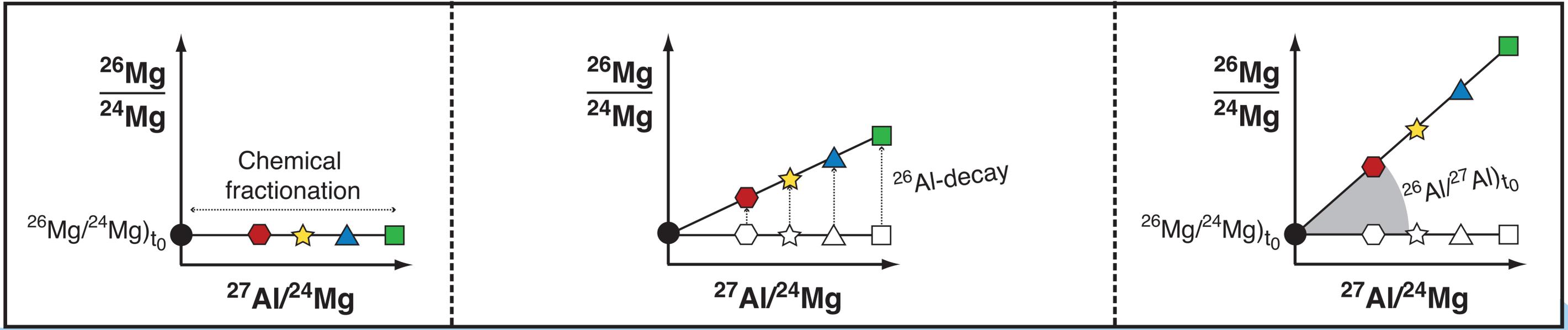
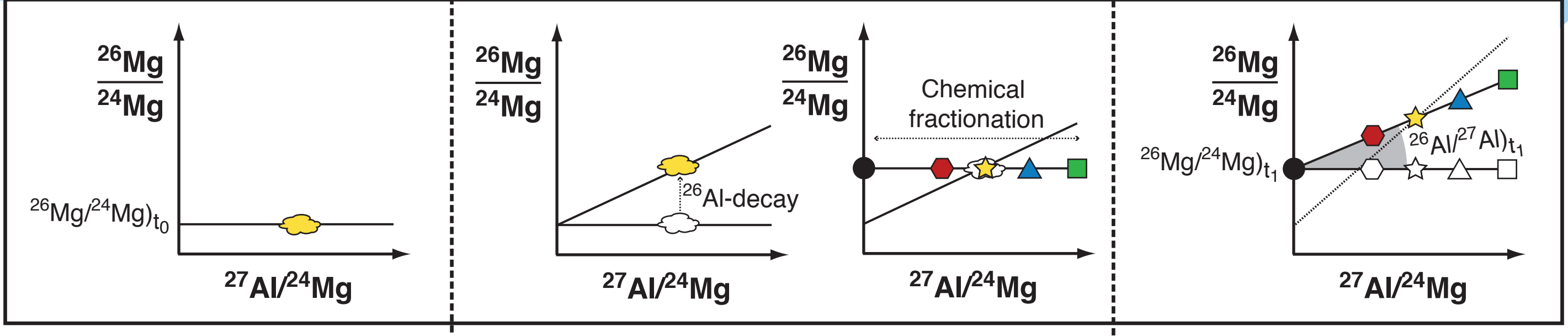

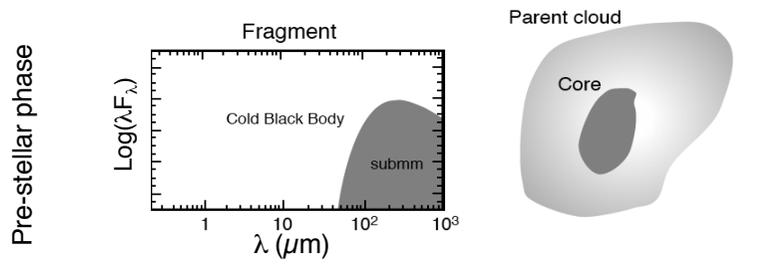
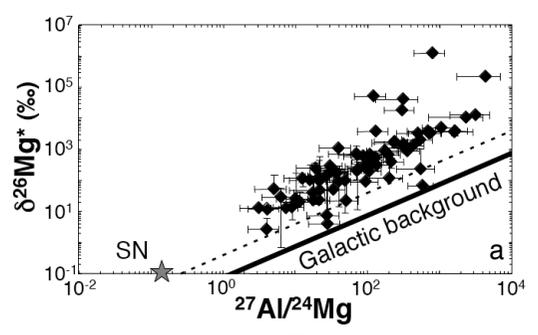
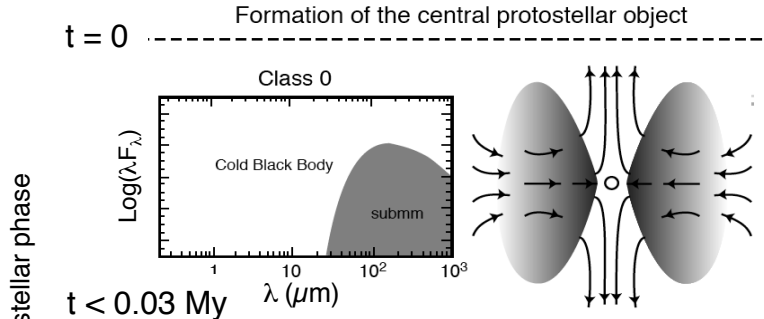
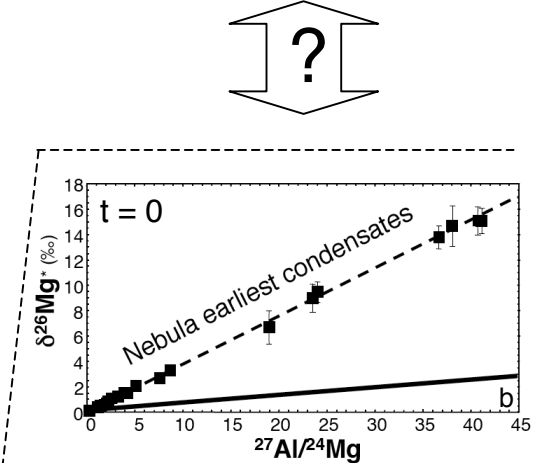
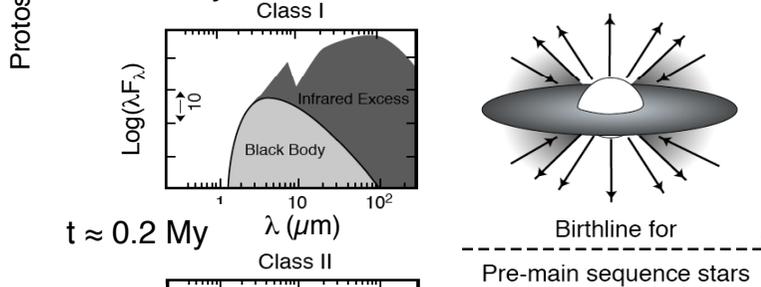
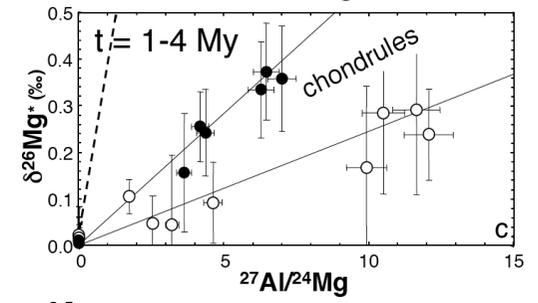
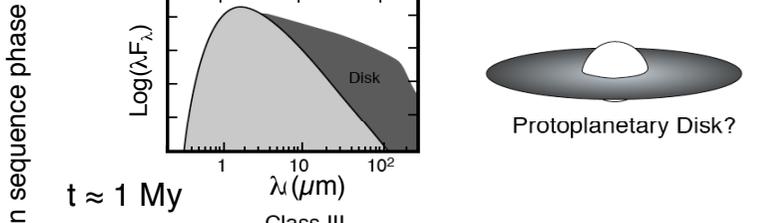
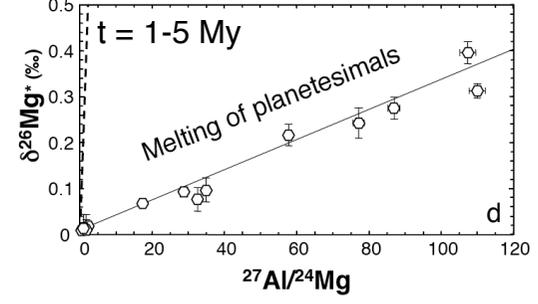
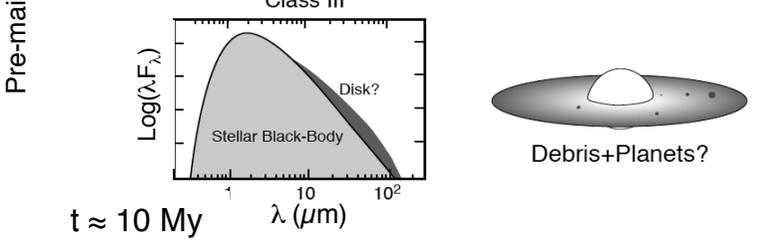
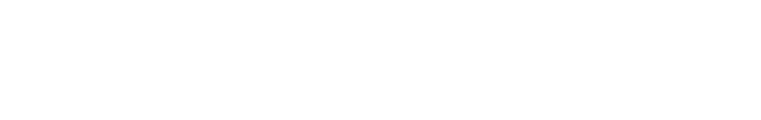

## Condensation of 1-10 μm dust from nebular gas

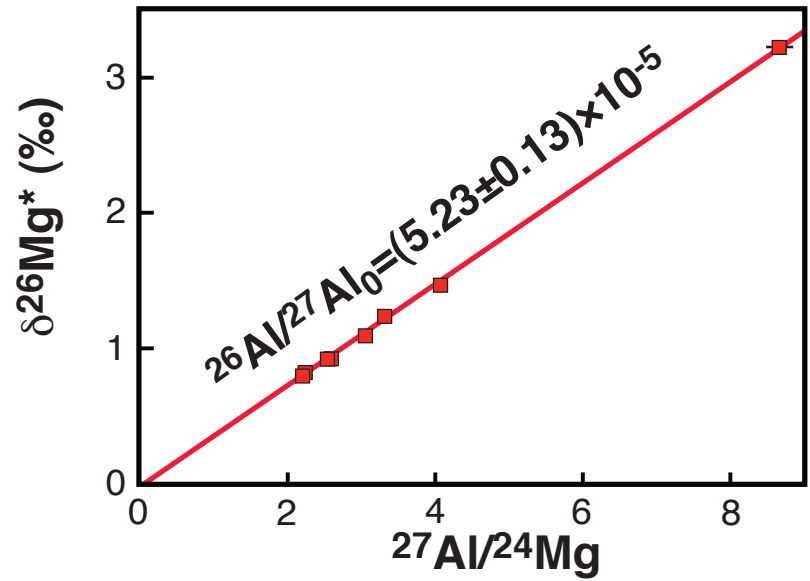

Bulk isochron of CAIs
(Jacobsen et al., 2008)

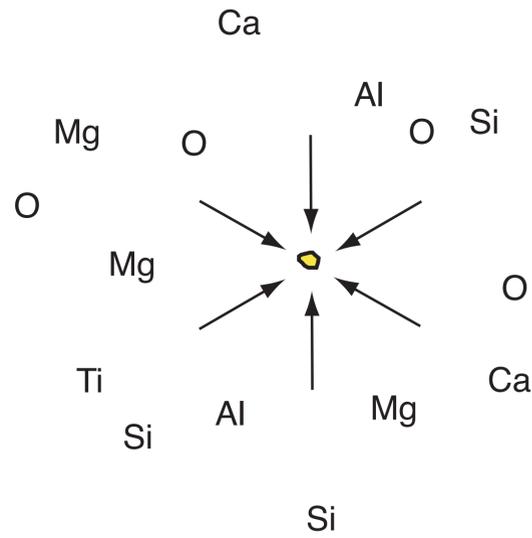

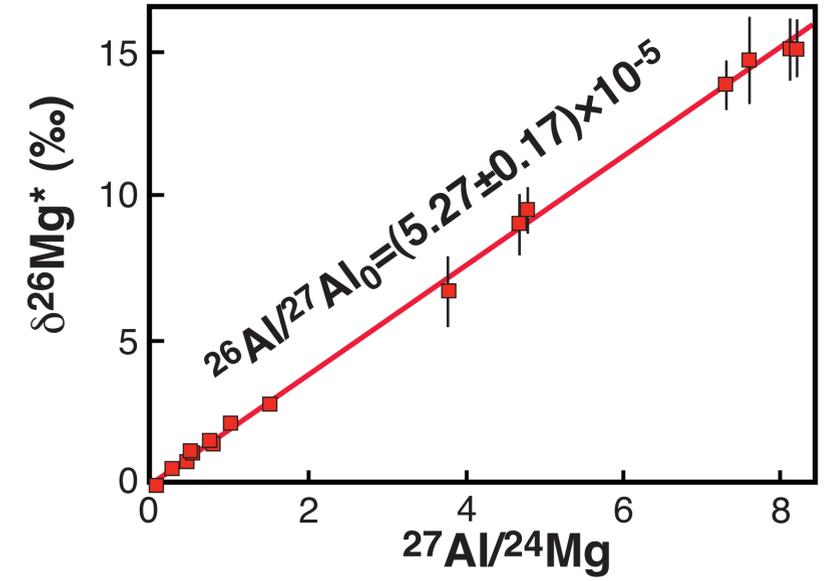

Internal isochron of a fine-grained CAI
(MacPherson et al., 2010a)

## Agglomeration into mm-cm objects

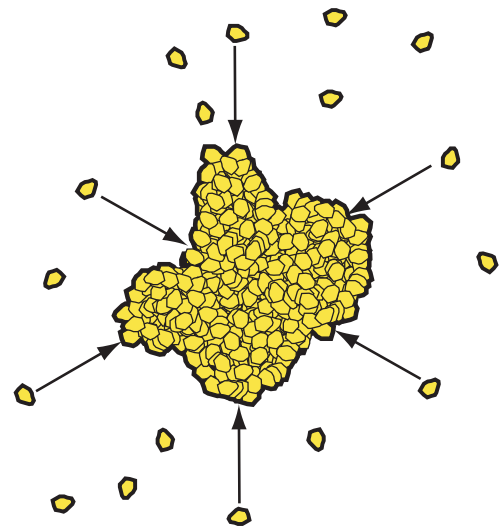

**< 50 ky**

## Melting and crystallization

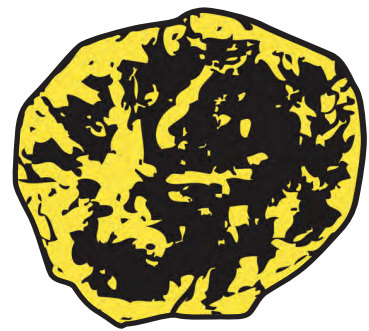

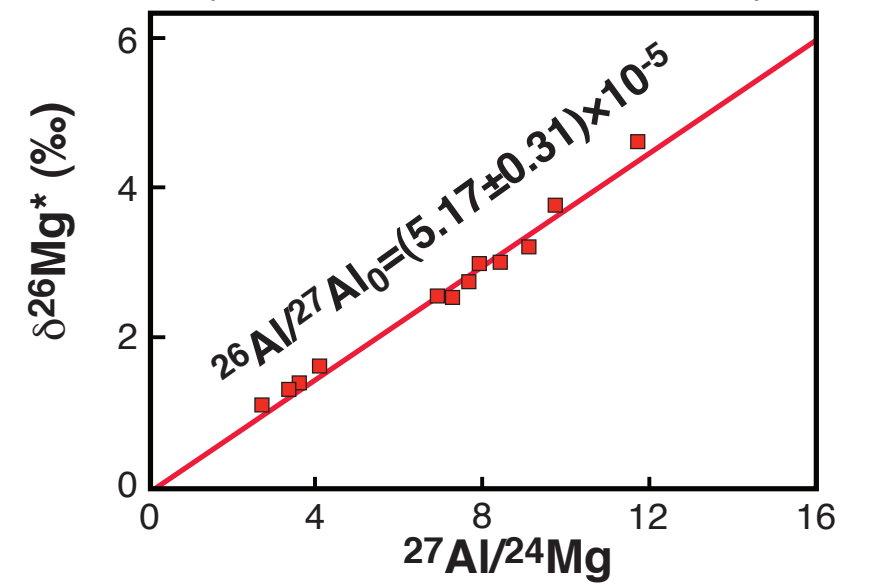

Internal isochron of an igneous CAI
(MacPherson et al., 2010b)

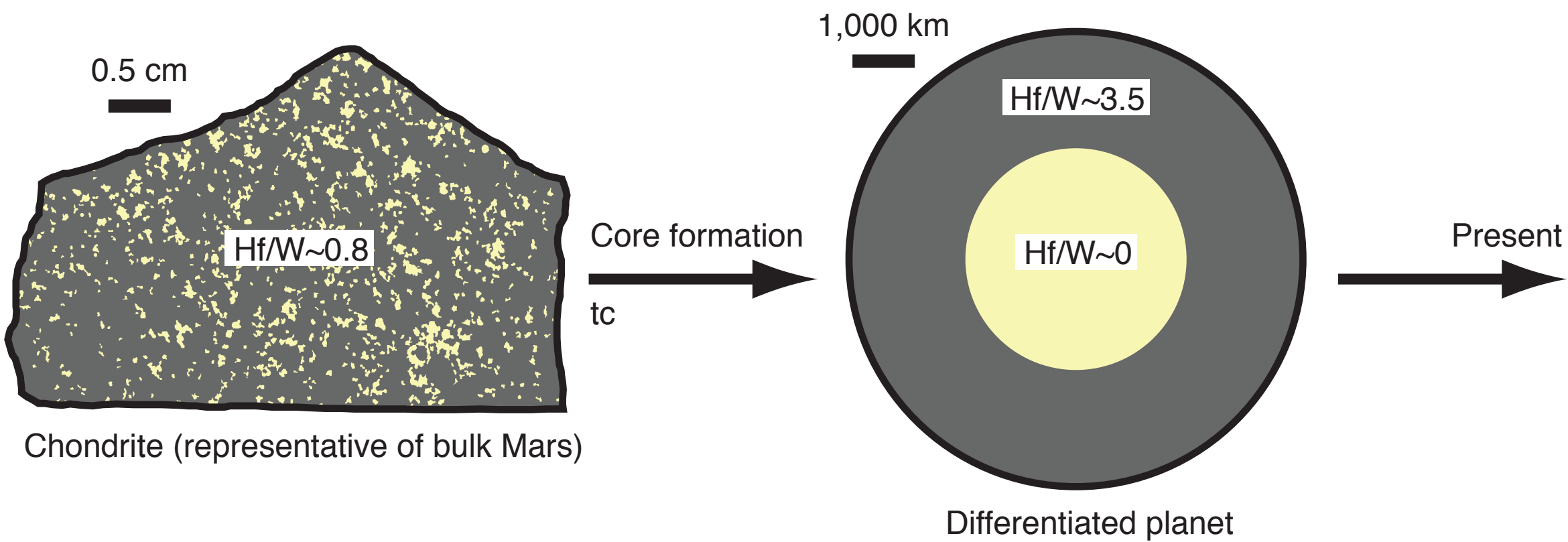
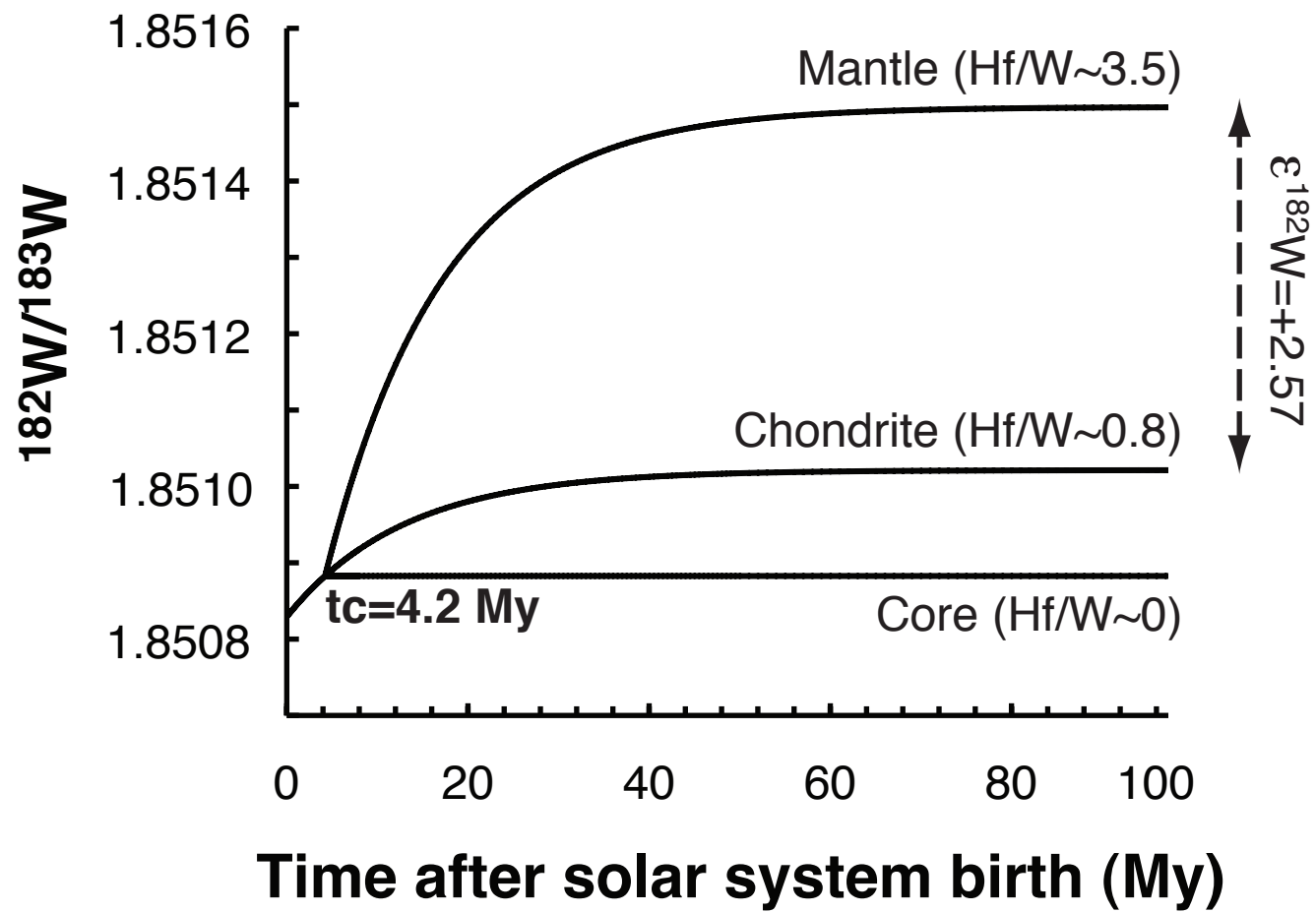

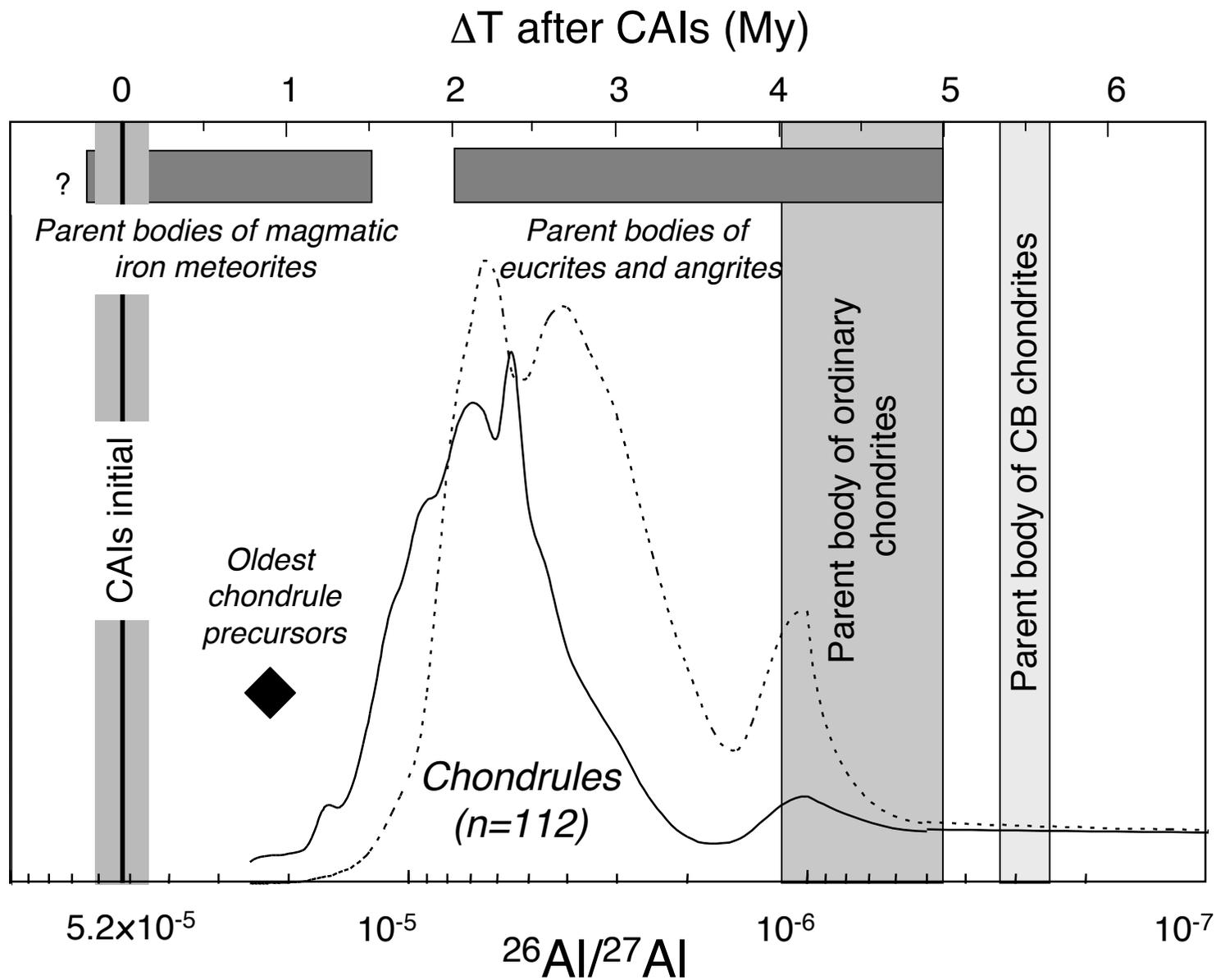

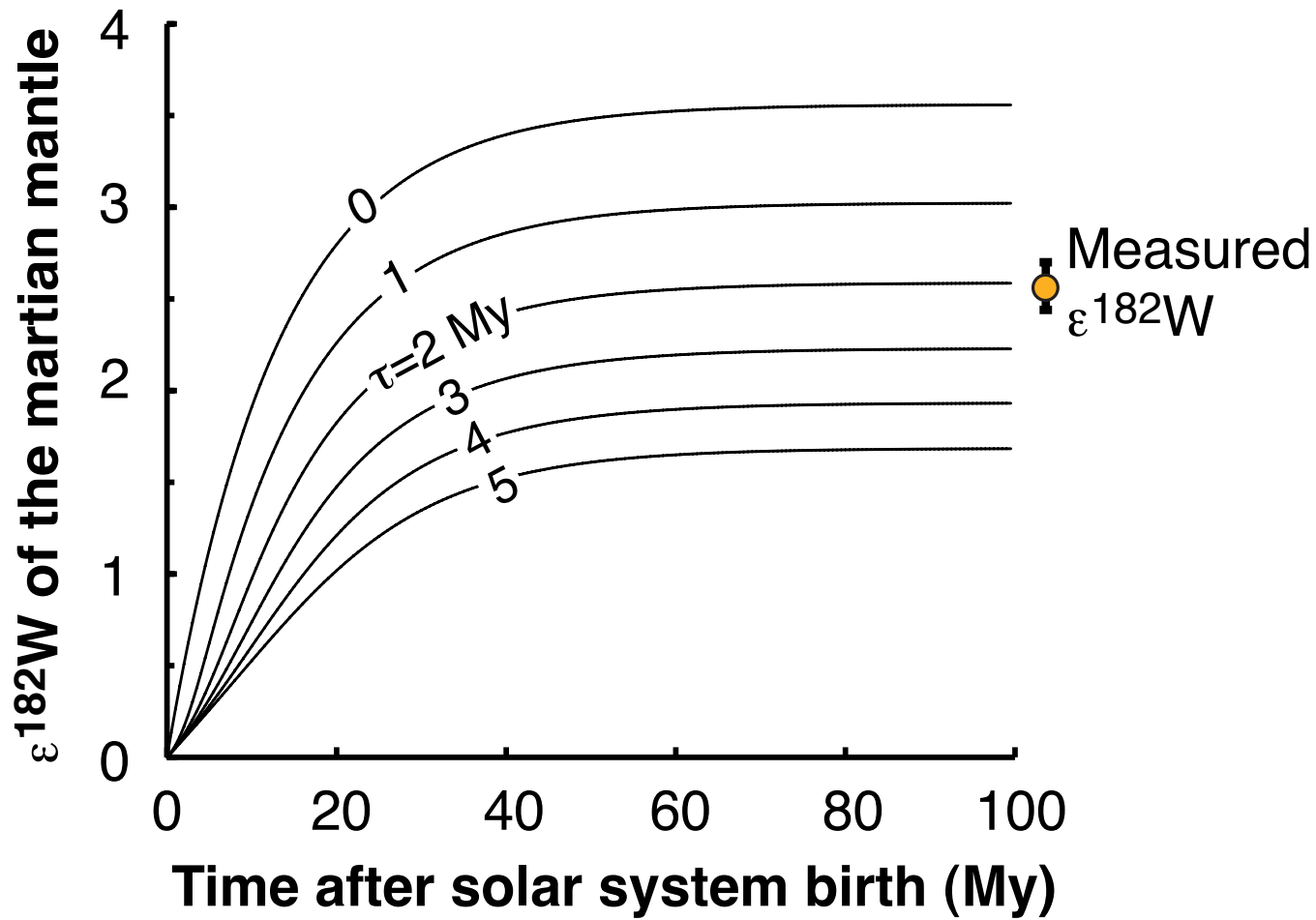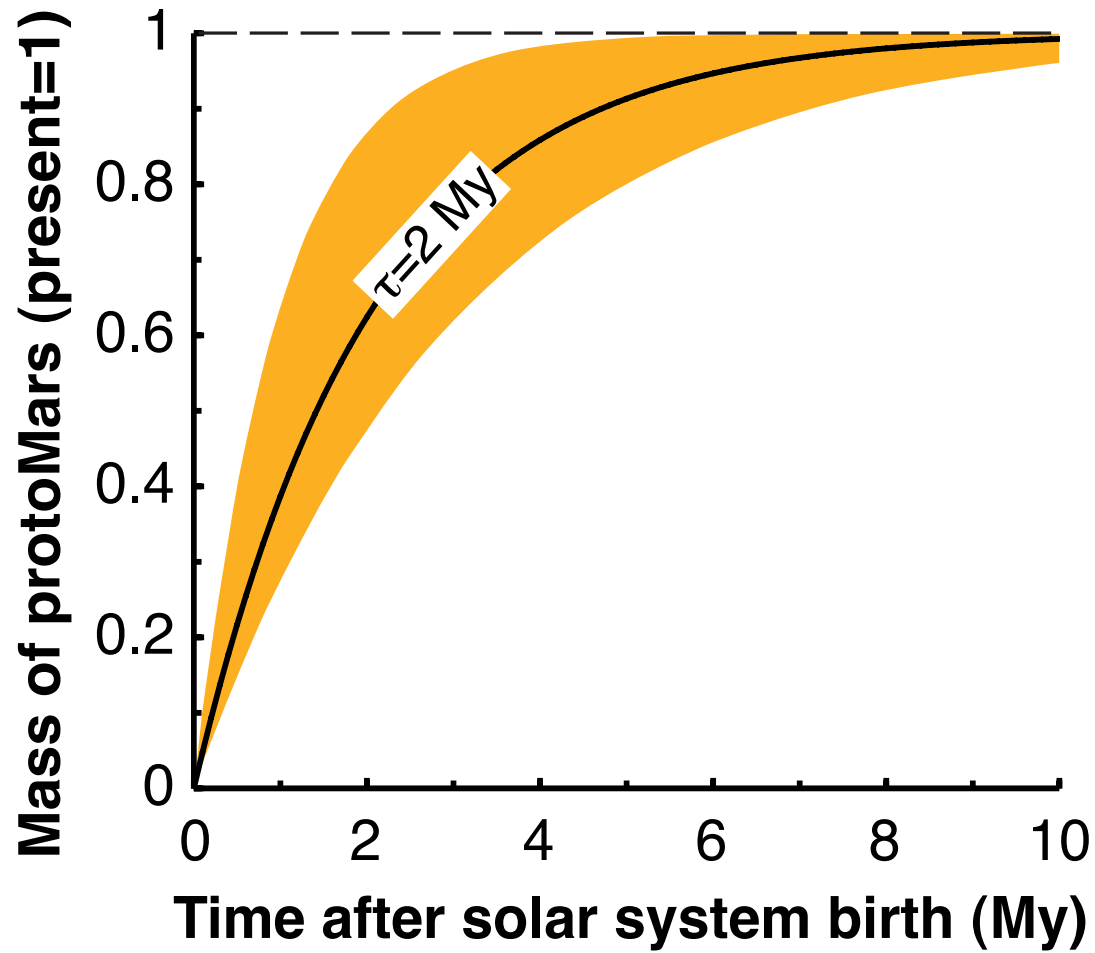

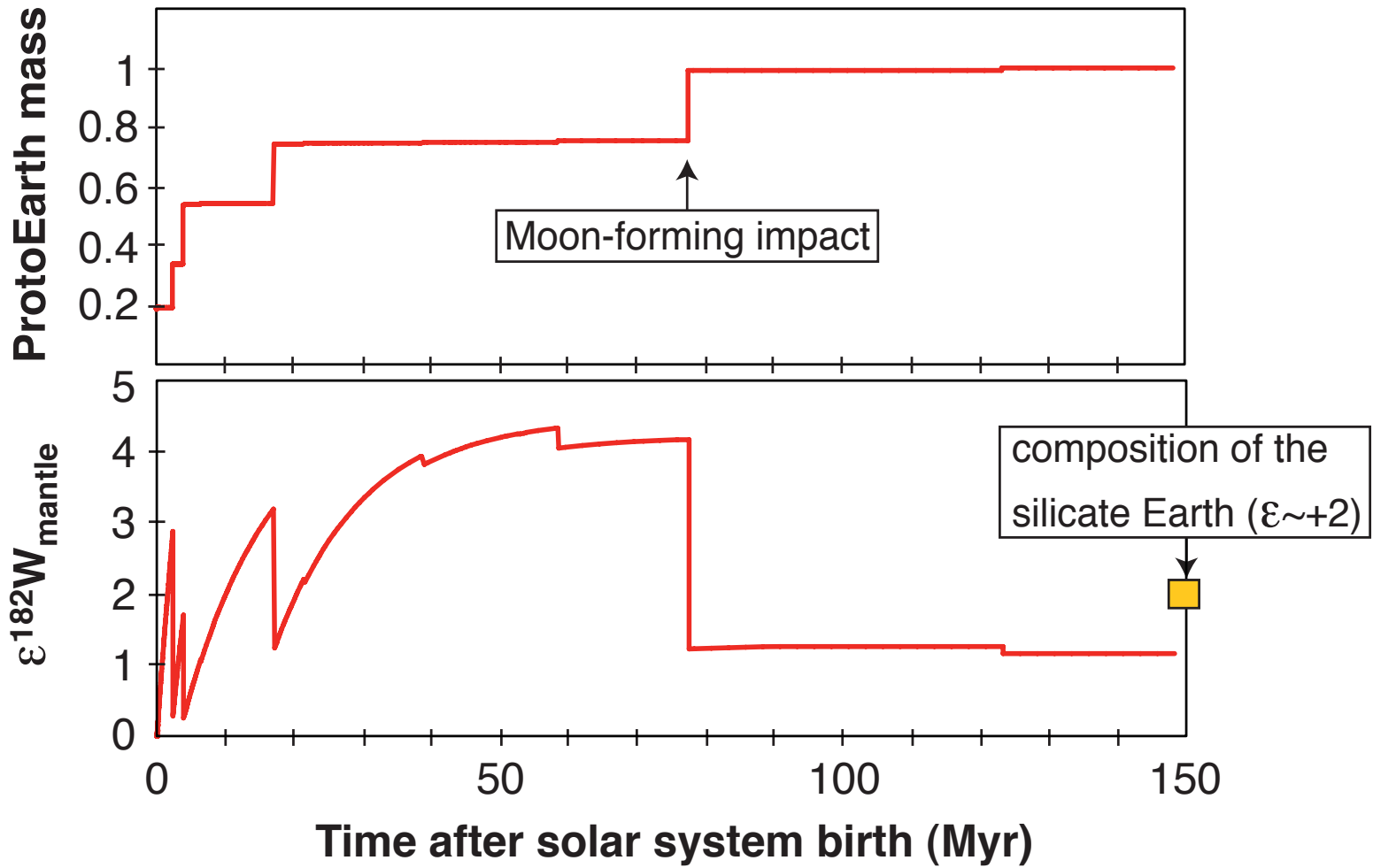

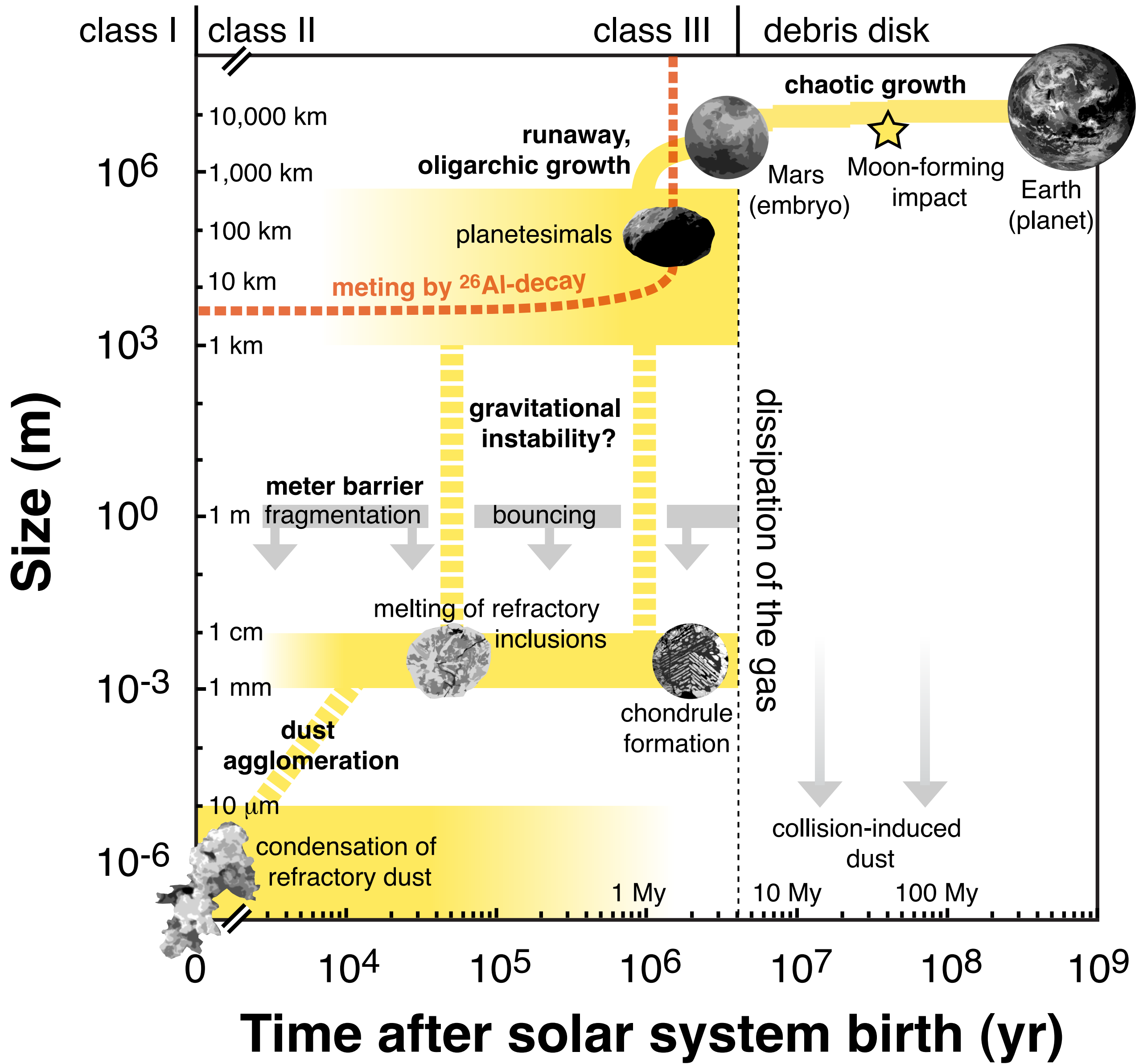

**Table 1.** Extinct radioactivities in meteorites

| Parent nuclide | Half-life (My) | Decay constant (My$^{-1}$) | Daughter nuclide | Estimated initial abundance | Note |
|---|---|---|---|---|---|
| Samarium-146 | 103 | 0.00673 | Neodymium-142 | $^{146}$Sm/$^{144}$Sm=(8.4±0.5)×10$^{-3}$ | ★★★ |
| Plutonium-244 | 80 | 0.0087 | Fission products | $^{244}$Pu/$^{238}$U=(6.8±1.0)×10$^{-3}$ | ★★★ |
| Iodine-129 | 15.7 | 0.0441 | Xenon-129 | $^{129}$I/$^{127}$I=(1.19±0.20)×10$^{-4}$ | ★★★ |
| Hafnium-182 | 8.9 | 0.078 | Tungsten-182 | $^{182}$Hf/$^{180}$Hf=(9.72±0.44)×10$^{-5}$ | ★★★ |
| Manganese-53 | 3.74 | 0.185 | Chromium-53 | $^{53}$Mn/$^{55}$Mn=(6.28±0.66)×10$^{-6}$ | ★★★ |
| Beryllium-10 | 1.385 | 0.500 | Boron-10 | $^{10}$Be/$^{9}$Be=(7.0±0.8)×10$^{-4}$ | ★★★ |
| Aluminum-26 | 0.717 | 0.967 | Magnesium-26 | $^{26}$Al/$^{27}$Al=(5.23±0.13)×10$^{-5}$ | ★★★ |
| Niobium-92 | 34.7 | 0.0200 | Zirconium-92 | $^{92}$Nb/$^{93}$Nb=(1.6±0.3)×10$^{-5}$ | ★★★ |
| Palladium-107 | 6.5 | 0.11 | Silver-107 | $^{107}$Pd/$^{108}$Pd=(5.9±2.2)×10$^{-5}$ | ★★ |
| Iron-60 | 2.62 | 0.265 | Nickel-60 | (7.9±2.8)×10$^{-9}$<$^{60}$Fe/$^{56}$Fe<(6.3±2.0)×10$^{-7}$ | ★★ |
| Chlorine-36 | 0.301 | 2.30 | Sulfur-36 (1.9 %), argon-36 (98.1 %) | $^{36}$Cl/$^{35}$Cl>(17.2±2.5)×10$^{-6}$ | ★★ |
| Curium-247 | 15.6 | 0.0444 | Uranium-235 | $^{247}$Cm/$^{238}$U=(5.5±2.0)×10$^{-5}$ | ★ |
| Lead-205 | 15.1 | 0.0459 | Thallium-205 | $^{205}$Pb/$^{204}$Pb=(1.0±0.4)×10$^{-3}$ | ★ |
| Cesium-135 | 2.3 | 0.30 | Barium-135 | $^{135}$Cs/$^{133}$Cs=(4.8±0.8)×10$^{-4}$ | ★ |
| Calcium-41 | 0.102 | 6.80 | Potassium-41 | $^{41}$Ca/$^{40}$Ca=(1.41±0.14)×10$^{-8}$ | ★ |
| Beryllium-7 | 1.46×10$^{-7}$ | 6.86×10$^{6}$ | Lithium-7 | $^{7}$Be/$^{9}$Be=0.0061±0.0013 | ★ |
| Technetium-97 | 4.21 | 0.16464 | Molybdenum-97 | $^{97}$Tc/$^{92}$Mo<3×10$^{-6}$ | < |
| Technetium-98 | 4.2 | 0.17 | Ruthenium-98 | $^{98}$Tc/$^{96}$Ru<2×10$^{-5}$ | < |
| Tin-126 | 0.23 | 3.0 | Tellurium-126 | $^{126}$Sn/$^{124}$Sn<7.7×10$^{-5}$ | < |